\documentclass[final,3p,times,authoryear]{elsarticle}

\usepackage{amssymb}
\usepackage{amsthm}
\usepackage{graphics} 
\usepackage{epsfig} 
\usepackage{amsmath} 
\usepackage{hyperref}

\usepackage{graphicx}
\usepackage{subfigure}

\journal{Transportation Research Part C: Emerging Technologies}

\usepackage{amssymb}

\usepackage[figuresright]{rotating}

\begin{document}

\begin{frontmatter}




\title{Modelling supported driving as an optimal control cycle: Framework and model characteristics}

\author[a]{\large Meng Wang}
\author[b]{Martin Treiber}
\author[a]{Winnie Daamen}
\author[a]{Serge P. Hoogendoorn}
\author[a]{Bart van Arem}
\address[a]{Delft University of Technology, Faculty of Civil Engineering and Geosciences, Stevinweg 1, 2600 GA, Delft, the Netherlands.}
\address[b]{Dresden University of Technology, Institute for Transport \& Economics, Wurzburger Str. 35, 01062 Dresden, Germany.}



\begin{abstract}
Driver assistance systems support drivers in operating vehicles in a safe, comfortable and efficient way, and thus may induce changes in traffic flow characteristics. This paper puts forward a receding horizon control framework to model driver assistance and cooperative systems. The accelerations of automated vehicles are controlled to optimise a cost function, assuming other vehicles driving at stationary conditions over a prediction horizon. The flexibility of the framework is demonstrated with controller design of Adaptive Cruise Control (ACC) and Cooperative ACC (C-ACC) systems. The proposed ACC and C-ACC model characteristics are investigated analytically, with focus on equilibrium solutions and stability properties. The proposed ACC model produces plausible human car-following behaviour and is unconditionally locally stable. By careful tuning of parameters, the ACC model generates similar stability characteristics as human driver models. The proposed C-ACC model results in convective downstream and absolute string instability, but not convective upstream string instability observed in human-driven traffic and in the ACC model. The control framework and analytical results provide insights into the influences of ACC and C-ACC systems on traffic flow operations.
\end{abstract}


\end{frontmatter}


\section{Introduction}
Advanced Driver Assistance Systems (ADAS) aim to support drivers or take over the driving tasks to operate vehicles in a safe, comfortable and efficient way \citep{Varaiya1991}. This includes cooperative systems, where equipped vehicles are connected to and collaborate with each other through Vehicle-to-Vehicle (V2V) or Vehicle-to-Infrastructure (V2I) communications \citep{Williams1992}. Considerable efforts have been dedicated to ADAS control design and investigation of the resulting traffic flow properties. Among them, Adaptive Cruise Control (ACC) systems attract most of the attention due to the early availability in the market. The most widely reported ACC model is a proportional derivative (PD) controller, where the vehicle acceleration is proportional to the gap (net distance headway) and relative speed with respect to the preceding vehicle (derivative of gap) at car-following conditions. This controller has been well examined \citep{Swaroop1994,Godbole1999,VanderWerf2002}, and is essentially a Helly car-following model \citep{Helly1959}. Extensions of this controller class have been reported to include acceleration of the predecessor \citep{VanderWerf2002,VanArem2006} or multi-anticipative behaviour \citep{Wilmink2007} in the controller. However, there is no safety mechanism in this model. Under critical conditions, ACC systems have to be overruled by drivers and hard braking has to be performed to avoid collision \citep{Godbole1999}. Some researchers \citep{Hasebe2003} used the Optimal Velocity Model (OVM) to describe the controlled vehicle behaviour and proposed a cooperative driving system under which the desired speed is determined not only by the gap to the vehicle in front but also by the gap to the vehicle behind. Unfortunately, the optimal velocity model is not collision free under realistic parameters \citep{Treiber2000}. The Intelligent Driver Model (IDM) is used to design ACC controllers with a driving strategy that varies parameters according to traffic situations to mitigate congestion at bottlenecks \citep{Kesting2008c,Treiber2010a}. Other controllers are reported by \cite{Swaroop1994} and \cite{Ioannou1993}.  The resulting traffic flow characteristics of ADAS differ among the controller and parameter settings. The increase of capacity is mainly a result of shorter time headways compared to human drivers \citep{Rao1993,Kesting2008c}, while choosing a larger time headway could cause negative impacts on capacity \citep{Minderhoud1999,VanderWerf2002}. 
Regarding the stability, some authors provide evidence that ACC/CACC systems improve flow stability \citep{Hasebe2003,Davis2004,VanArem2006,Naus2010}, while others  \citep{Marsden2001} are more conservative on the stabilisation effects of ACC systems.

ADAS and Cooperative Systems have a direct influence on the vehicular behaviour and consequently on flow operations. The lack of clarity on aggregated impacts of ADAS in literature calls for new insights into the model properties of ADAS and cooperative systems. Furthermore, the increasing public concerns on traffic congestion and environment stimulate the need for development of driver assistance systems that can fulfil multiple objectives, cooperate with each other and operate vehicles in an optimal way. It is however difficult to use the existing phenomenological ADAS controllers to achieve all these objectives.

This contribution generalises previous work on driver behaviour \citep{Hoogendoorn2009a} to a control framework for driver assistance and cooperative systems. The framework is generic in such a way that different control objectives, i.e. safety, comfort, efficiency and sustainability, can be optimised. It is assumed that accelerations of ADAS vehicles are controlled to optimise a cost function reflecting multiple control objectives. Under the framework, we propose a complete ACC controller, which produces plausible human car-following behaviour at both microscopic and macroscopic level. The controller can be applied to all traffic situations, i.e. not only car-following and free driving conditions, but also safety-critical conditions such as approaching standstill vehicles with high speeds. The flexibility in the system and cost specification allows modelling a Cooperative ACC (C-ACC) controller, where an equipped vehicle exhibits cooperative behaviour by optimising the joint cost of both itself and its follower.

The aggregated flow characteristics of the ACC/C-ACC models are investigated analytically, with a focus on equilibrium solutions and (linear) stability analysis. Analytical criteria to quantify the influence on the model stability  due to cooperative behaviour are derived.

The rest of the paper is structured as follows. Section 2 presents the modelling framework and solution approach, with several examples showing the application of the framework. Section 3 gives the analytical solutions at equilibrium conditions, criteria for string stability and the method for classification of string instability types. Section 4 gives insights into the model characteristics of the example controllers. Conclusions and future work are discussed in section 5.

\section{Control framework for supported driving}
In this section, we first present the underlying assumptions and mathematical formulation of the control framework. The optimal control problem is solved using the dynamic programming approach, and the framework is applied to design ACC and cooperative ACC controllers.

\subsection{Design assumptions and control objectives}
The controller framework is based on the following assumptions:
\begin{enumerate}
\item A controlled vehicle adapts its speed or changes lanes to minimise a certain cost function, reflecting the control objectives.
\item A controlled vehicle has all information regarding (relative) positions and speeds of other vehicles influencing its control decisions.
\item Other vehicles influencing the control decisions are driving at stationary conditions within the prediction horizon, i.e. accelerations equal zero.
\item Control decisions are updated at regular time intervals.
\item Longitudinal manoeuvres of ADAS equipped vehicles are under automated control.
\end{enumerate}
For the sake of analytical tractability, we only consider deterministic cases without time delay in this contribution, i.e. there is no noise in the information regarding other vehicles and the control decisions can be executed immediately. The control framework is generic in that it allows one to include stochastic processes and time lags in the controller \citep{Meng2012c}.

Control decisions are made to fulfil some control objectives, which can be a subset of the following:
\begin{enumerate}
\item To maximise travel efficiency;
\item To minimise lane-changing manoeuvres;
\item To minimise risk;
\item To minimise fuel consumption and emissions;
\item To maximise smoothness and comfort.
\end{enumerate}
The importance of each of these objectives can vary according to design preferences, traffic conditions, or individual vehicles, e.g. some systems may give priority to safe driving, while others prefer travel efficiency, accepting smaller headways and higher risk if other influencing factors (speed and relative speed) are kept constant.

\subsection{Supported driving as a receding control problem}
The proposed framework formulates the movements of ADAS equipped vehicles as a receding horizon control (also referred to as model predictive control) process, which entails solving an optimal control problem subject to system dynamics and other constraints on system state and control input \citep{Hoogendoorn2009a}. Fig. 1 shows the schematic graph of the receding horizon control process. 
At time instant $t_k$, the controller of equipped vehicle $n$ receives the positions and speeds of other vehicles from (erroneous) observations either made by its on-board sensors or transmitted from other sensors through V2V and/or V2I communication. Based on this information and past state, the controller estimates the current state of the system $\textbf{x}$, and uses a (system dynamics) model to predict the future state of the system in a time horizon $T_p$, with the estimate of the system state at $t_k$ as the initial condition. The control input $\textbf{u}$, i.e. acceleration or lane choice, is determined to minimise the cost $J$ accumulated in the prediction horizon reflecting, for instance, deviation of the future state from the desired state. The on-board actuators will execute the control input $\textbf{u}$ at time $t_k$. As the vehicle manoeuvres, the system changes, and the optimal control signal $\textbf{u}$ will be recalculated with the newest information regarding the system state at regular time intervals, i.e. at time $t_{k+1} = t_k + \Delta t$.

\begin{figure}
\centering
\includegraphics[height= 5 cm]{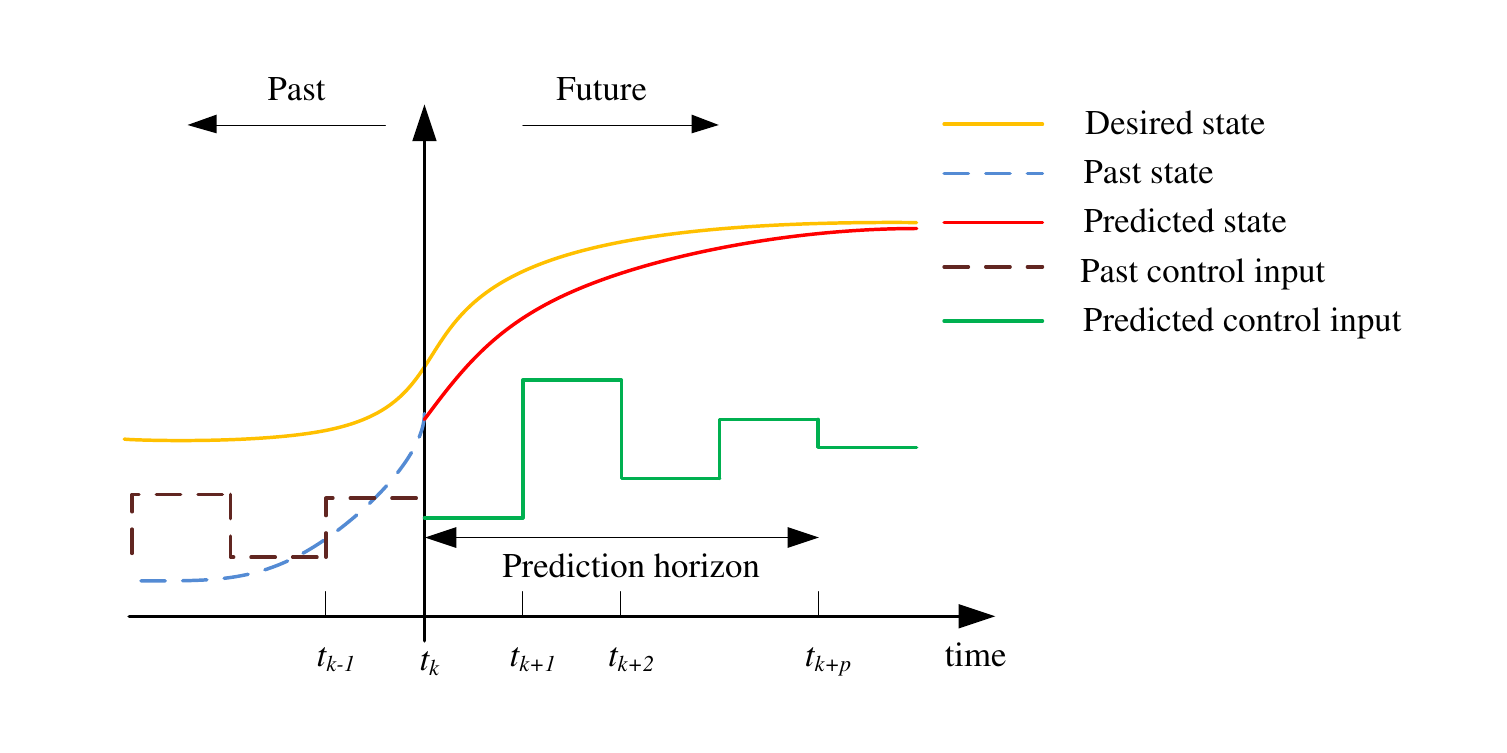}
\caption{Scheme of receding horizon control.}.
\label{fig:ISTTT_MPC}
\end{figure}

\subsection{Mathematical formulation of longitudinal control}
\subsubsection{State prediction model}
The system state $\textbf{x}$ from the perspective of ACC vehicle $n$ is fully described by the gap (net distance headway) $s$, the relative speed $\Delta v$ with respect to its predecessor and its own speed $v$, where $\textbf{x}=(x_1,x_2, x_3)^T=(s_n, \Delta v_n,v_n)^T$ with $\Delta v_n = v_{n-1} - v_n$. The system dynamics follow the deterministic kinematic equations:
\begin{equation}
\label{eq:systemdynamics}
{\mathrm{d} \over \mathrm{d}t} \textbf{x} ={\mathrm{d} \over \mathrm{d}t} \left(\begin{array}{c}  s_n \\\Delta v_n \\ v_n\end{array}\right) =  \left(\begin{array}{c} \Delta v_n \\ u_{n-1} - u_n \\ u_n \end{array}\right)= \textbf{f}(\textbf{x},\textbf{u})
\end{equation}
where $u_n$ denotes the acceleration of vehicle $n$, which is the control input in this model. $u_{n-1}$ denotes the acceleration of the predecessor, which equals zero within the prediction horizon based on our assumption. The considered system is a \textit{time invariant system}, i.e. the system dynamics model \textbf{f} does not depend explicitly on time $t$. 

Notice that when applying the controller, other vehicles may not travel at constant speed, which implies a mismatch between the prediction model and the system due to the constant-speed heuristic. The feedback nature of the receding horizon process, which entails reassessing the control input  at regular time intervals $\Delta t$ with the newest information of other vehicles, is permanently corrected, and thus robust to the mismatch.

For Cooperative ACC (C-ACC) controllers, the system state for vehicle $n$ is extended to include the situation of its follower $n+1$, $\textbf{x} = (s_n, \Delta v_n,v_n, s_{n+1}, \Delta v_{n+1},v_{n+1})^T$, where $s_{n+1}, \Delta v_{n+1}$ and $v_{n+1}$ denote the gap, relative speed and speed of the follower of the controlled vehicle respectively. The system dynamics now follow:
\begin{equation}
\label{eq:systemdynamics_cooperative}
{d \over dt} \textbf{x} ={d \over dt} \left(\begin{array}{c}  s \\\Delta v_n \\ v_n \\s_{n+1} \\\Delta v_{n+1} \\ v_{n+1} \end{array}\right) =  \left(\begin{array}{c} \Delta v_n \\  u_{n-1} - u_n \\ u_n \\ \Delta v_{n+1} \\ u - u_v \\ u_{n+1} \end{array}\right)= \textbf{f}(\textbf{x},\textbf{u})
\end{equation}
with $u_{n+1}$ denoting the acceleration of the follower.  $u_{n-1}$ and $u_{n+1}$ equal zero within the prediction horizon.

\subsubsection{Cost formulation}
We formulate the cost of car following, given that the control input $\textbf{u} = \{u_n(\tau) | \tau \ge t_k \}$ is applied , using the following functional:

\begin{equation}
\label{eq:costfunctional}
J(t_k, \textbf{x} | \textbf{u}) = \int_{t_k}^{t_k + T_p} e^{-\eta \tau} {\cal L}
(\textbf{x}, \textbf{u},\tau) d\tau + e^{-\eta (t_k+T_p)} \phi(\textbf{x}(t_k + T_p))
\end{equation}
with $T_p$ denoting the prediction horizon. The cost functional $J(t_k, \textbf{x} | \textbf{u})$ describes the expected cost (or disutility) given the current state of the system $\textbf{x}(t_k)$, the control input $\textbf{u}$ and the evolution of the system, starting from the current time $t_k$ to terminal time $t_k + T_p$.  In Eq. (\ref{eq:costfunctional}), $\cal{L}$ denotes the so-called \textit{running cost}, describing the cost incurred during an infinitesimal period $[\tau,\tau+d\tau)$, which are additive over time. $ \phi $ denotes the so-called \textit{terminal cost}, which reflects the cost remaining at the terminal time.

The parameter $\eta \geq 0$ with a unit of $s^{-1}$ denotes the so-called discount factor \citep{Fleming1993}, which reflects some trade-off between cost incurred in the near term and future cost. $\eta = 0$ implies that the controller weighs the future cost similar to the current cost, which may be the case if the controller can predict the dynamics of the predecessor behaviour fairly well. $\eta>>0$ results in a short-sighted driving behaviour where the controller optimises the immediate situation and does not care too much about the future. Particularly, the cost after a future horizon $[0, \frac{1}{\eta})$ decreases exponentially. 

Notice that if $\eta = 0$ and $T_p<\infty$, the considered problem pertains to a finite horizon optimal control problem with un-discounted cost \citep[e.g.,][]{Fleming1993}. Solving this type of problem entails choosing a terminal cost $\phi$ to ensure expected controller behaviour and computational feasibility, which is not trivial \citep{Chen1998}.  An alternative is to set $\eta > 0$ and $T_p =\infty$, thus the weight for the terminal cost $ e^{-\eta T_p} $ equals zero. This removes the parameter  $T_p$ and relieves us from defining a terminal cost $\phi$. The considered problem becomes an infinite horizon optimal control problem with discounted cost \citep[e.g.,][]{Fleming1993}.

In the present work, we choose the infinite horizon problem with discounted cost. The optimal control problem is now described by the following mathematical program:
\begin{equation}
\label{eq:min}
\textbf{u}_{[t_k,\infty)}^* = \arg \min J(t_k, \textbf{x} | \textbf{u}) = \arg \min \int_{t_k}^{\infty} e^{-\eta \tau}{\cal L}(\textbf{x}, \textbf{u}) d\tau
\end{equation}
subject to:
\begin{equation}
\label{eq:state}
{\mathrm{d} \over \mathrm{d}t} \textbf{x} = f(\textbf{x},\textbf{u})
\end{equation}
The control input $\textbf{u}$ will be re-assessed at regular time intervals $\Delta t = t_{k+1} - t_k$ using the most current observations or estimates of the system state (at time $t_{k+1}$). 

Notice that in this contribution we consider multiple criteria for the optimisation, i.e. safety, efficiency, and comfort, but transform the supported driving task into a single-objective mathematical optimisation problem (Eqs (\ref{eq:min}, \ref{eq:state})) by assuming fixed weights for different criteria.

\subsection{Solution approach based on Dynamic Programming}
\label{sec:solutionapproach}
Here we briefly discuss the solution to the considered problem of Eqs. (\ref{eq:min}, \ref{eq:state}), based on the well-known dynamic programming approach.

Let us denote $W(t_k,\textbf{x})$ as the so-called $value function$, which is the optimal cost function under optimal control $\textbf{u}^*$: 
\begin{equation}
W(t_k,\textbf{x}) = J(t_k, \textbf{x} | \textbf{u}^*)
\end{equation}
 
Applying Bellman's Principle of Optimality yields the \textit{ Hamilton-Jacobi-Bellman (HJB)} equation with discount factor as \citep{Fleming1993}:
\begin{equation}
\label{eq:HJB}
\eta W(\textbf{x}) = {\cal H} \left( \textbf{x},\textbf{u}^*,\frac{\partial W(\textbf{x})}{\partial \textbf{x}} \right)
\end{equation}
where  $\cal{H}$ is the so-called Hamilton equation (Hamiltonian), which satisfies:
\begin{equation}
\label{eq:Hamilton}
{\cal H} \left( \textbf{x},\textbf{u}^*,\frac{\partial W(\textbf{x})}{\partial \textbf{x}} \right)=\underset{\textbf{u}}{\min} \left({\cal L} + \frac{\partial W(\textbf{x})}{\partial \textbf{x}} \cdot \textbf{f} \right)
\end{equation}

Let  $ \boldsymbol{\lambda} =  \frac{\partial W(\textbf{x})}{\partial \textbf{x}}$ denote the so-called \textit{co-state} or \textit{marginal cost} of the state $\textbf{x}$, reflecting the relative extra cost of $W$ due to making a small change  $\delta \textbf{x}$ on the state $\textbf{x}$. Taking the partial derivative of Eq. (\ref{eq:HJB}) with respect to state $\textbf{x}$ gives:
\begin{equation}
\label{eq:costatesolution}
 \boldsymbol{\lambda} =  \frac{1}{\eta} \frac{\partial  \cal H}{\partial \textbf{x}} = \frac{1}{\eta} \frac{\partial  \cal{L}}{\partial \textbf{x}} + \frac{1}{\eta}\frac{\partial ( \boldsymbol{\lambda} \cdot \textbf{f}) }{\partial \textbf{x}} 
\end{equation}

Using the Hamiltonian of Eq. (\ref{eq:Hamilton}), we can derive the following necessary condition for the optimal control $\textbf{u}^*$:
\begin{equation}
\label{eq:necessary_condition}
{\cal H} (\textbf{x},\textbf{u}^*,\boldsymbol{\lambda}) \le {\cal H}(\textbf{x},\textbf{u},\boldsymbol{\lambda}), \hspace{0.25cm} \forall \textbf{u}
\end{equation}
In nearly all cases, this requirement will enable expressing the optimal  control $\textbf{u}^*$ as a function of the state $\textbf{x}$ and the co-state $\boldsymbol{\lambda}$.

Taking the necessary condition of $\frac{\partial \cal H}{\partial \textbf{u}}=0$ gives the following optimal control law for ACC vehicle $n$:
\begin{equation}
\label{eq:optimalcontrol}
 \textbf{u}^* = \lambda^{\Delta v_n} -\lambda^{v_n}  
\end{equation}
where $\lambda^{\Delta v_n}$ and $\lambda^{ v_n}$ denote the co-state of relative speed and the co-state of speed respectively, and are given by:
\begin{equation}
\label{eq:costate}
\lambda^{\Delta v_n} = \frac{1}{\eta}\frac{\partial {\cal L}}{\partial \Delta v_n} + \frac{1}{\eta^2} \frac{\partial {\cal L}}{\partial s_n}  \mbox{ , } \lambda^{v_n} =  \frac{1}{\eta}\frac{\partial {\cal L}}{\partial v_n} - \frac{1}{\eta^2} \frac{\partial {\cal L}}{\partial s_n}
\end{equation}
The optimal acceleration control law (\ref{eq:optimalcontrol}) states that the automated vehicle will increase its speed when the marginal cost of relative speed is larger than the marginal cost of speed, and decelerate when vice versa.

For the C-ACC controller, the change in the system state and dynamics results in the following optimal control law when applying the same solution approach:
\begin{equation}
\label{eq:optimalcontrol_cooperative}
\textbf{u}^* = \lambda^{\Delta v_n} - \lambda^{v_n} - \lambda^{\Delta v_{n+1}}
\end{equation}
with  $\lambda^{\Delta v_n}$ and $\lambda^{ v_n}$  given in (\ref{eq:costate}) and 
\begin{equation}
\lambda^{\Delta v_{n+1}} = \frac{1}{\eta} \frac{\partial {\cal L}}{\partial \Delta v_{n+1}} + \frac{1}{\eta^2} \frac{\partial {\cal L}}{\partial s_{n+1}}
\end{equation}
Equation (\ref{eq:optimalcontrol_cooperative}) shows that the optimal acceleration for a C-ACC vehicle is determined by the marginal costs of its relative speed and speed, as well as the marginal cost of the relative speed of its follower. Clearly, the inclusion of marginal cost of the follower's speed in the optimal control law captures the \textit{cooperative nature} of the C-ACC controller.

We emphasise that the control input $\textbf{u}$ is not limited to the control of a single vehicle. The framework allows simultaneous control of multiple vehicles, i.e. two controlled vehicles in a cooperative system.

\subsection{Example 1: ACC model}
As a first example, we present an ACC model that is collision-free and can generate plausible human driving behaviour using the proposed control framework.
\subsubsection{Cost specification and optimal acceleration}
We distinguish between cruising (free driving) mode and following mode for the proposed ACC system. In cruising mode, ACC vehicles try to travel at a user defined free speed $v_0$. In following mode, ACC vehicles try to maintain a gap-dependent desired speed $v_d$ while at the same time avoiding driving too close to the predecessor. For the sake of notation simplicity, we will drop the index $n$ in the ACC controller. Mathematically, the two-regime running cost function can be formulated as:
\begin{equation}
\label{eq:ACCcost}
{\cal L} =\left\{ \begin{array}{cl}
{ \underbrace{c_1 e^{\frac{s_0}{s}}\Delta v^2 \cdot\Theta(\Delta v)}_{safety} + \underbrace{c_2 (v_d(s) - v)^2}_{efficiency} + \underbrace{\frac{1}{2}u^2 }_{comfort} } &\mbox{ if } s \leq s_f = v_0\cdot t_d + s_0 \\
{\underbrace{c_3 (v_0 - v)^2}_{efficiency} + \underbrace{\frac{1}{2}u^2 }_{comfort}} 
&\mbox{ if } s > s_f = v_0\cdot t_d + s_0 \\
\end{array}  \right.
\end{equation}
where $s_f $ is the gap threshold to distinguish cruising mode ($s > s_f$) from following mode ($s\leq s_f$) and is calculated with $s_f = v_0\cdot t_d + s_0$, where $v_0$ is the free speed and $s_0$ is the distance between two cars at completely congested (standstill) conditions. $t_d$ denotes the user-defined desired time gap. $v_d(s)$ is the so-called \textit{desired speed} in following mode and is determined by : 
\begin{equation}
\label{eq:desiredspeed}
v_d(s) = \frac{s-s_0}{t_d}
\end{equation}
$\Theta$ is a delta function which follows the form:
\begin{equation}
\label{eq:theta}
\Theta(\Delta v)= \left\{ \begin{array}{cl}
{1} &\mbox { if $ \Delta v \leq 0$} \\
{0} &\mbox{ if $ \Delta v > 0 $} \\
       \end{array} \right.
\end{equation}
Equation (\ref{eq:ACCcost}) implies that the controller makes some trade-off among the safety cost, efficiency cost and comfort cost when following a preceding vehicle:
\begin{itemize}
\item The safety cost only incurs when approaching the preceding vehicle, i.e. $\Delta v<0$; $c_1 >0$ is a constant weight factor. The exponential term $e^{\frac{s_0}{s}}$ of the safety cost ensures a large penalty when driving too close to the predecessor, i.e. $s \leq s_0$. The safety cost is a monotonic decreasing function of gap $s$, reflecting the fact that the sensitivity to the relative speed tends to decrease with the increase of following distance. There is no safety cost in cruising mode. 
\item The efficiency cost term in following mode incurs deviating from the desired speed;  $c_2>0$ is a constant weight factor. The user-set desired time gap $t_d$ reflects driver preference and driving style, i.e. a smaller $t_d$ tends to an aggressive driving style, while a larger one means more timid driving behaviour. This cost also stems from the interaction with the predecessor, and  will not appear in the cruising mode.
\item The travel efficiency cost in cruising mode stems from not driving at free speed $v_0$, with a constant weight $c_3>0$.
\item The comfort cost is represented by penalising accelerating or decelerating behaviour. 
\end{itemize}

Employing the solution of Eq. (\ref{eq:optimalcontrol}) arrives at the following optimal control law:
\begin{equation}
\label{eq:optaACC}
u^* = \left\{ \begin{array}{cl}
{\frac{2 c_1e^{\frac{s_0}{s}}}{\eta} \left( \Delta v  - \frac{s_0\Delta v^2}{\eta s^2} \right)\cdot\Theta(\Delta v) +  \frac{2c_2}{\eta}\left( 1 + \frac{2}{\eta t_d} \right)\left(v_d(s) - v \right)} &\mbox { if $s \leq s_f$} \\
\frac{2c_3}{\eta}{(v_0-v)} &\mbox{ if $s > s_f$} \\
       \end{array} \right.    
\end{equation}
Equation (\ref{eq:optaACC}) shows that the optimal acceleration is a function of the state $\textbf{x} = (s, \Delta v, v)^T$. The first term in following mode (when $s\leq s_f$) describes the tendency to decelerate when approaching the predecessor, while the second term describes the tendency to accelerate when the vehicle speed is lower than the desired speed and the tendency to decelerate when vice versa. In cruising mode ACC vehicles adjust their speed towards the free speed  $v_0$ to minimise the efficiency cost, with an acceleration proportional to the speed difference with respect to the free speed.

In reality, the accelerations of vehicles are usually limited by the power train, i.e. $u \leq 2 m/s^2$. For the optimal acceleration function (\ref{eq:optaACC}), it achieves its maximum $u^*_{max,f}$ in following mode when $s = s_f$, $v = 0 km/h$, and $\Delta v\geq 0 km/h $ and achieves its maximum $u^*_{max,c}$ in cruising mode when $v=0 km/h$ for all $s > s_f$ and $\Delta v$: 
\begin{equation}
\label{eq:c2_max_a}
u^*_{max,f} = u(s_f,\Delta v,0) =\frac{2c_2v_0}{\eta}(1+\frac{2}{\eta t_d}) \mbox{  , for $\Delta v \geq 0$}
\end{equation}
and
\begin{equation}
u^*_{max,c}=a^*(s,\Delta v,0) = \frac{2c_3v_0}{\eta} \mbox{ ,  for $s>s_f$}
\end{equation}

To smooth the transition from following mode to cruising mode, we let $u^*_{max,f} =   u^*_{max,c}$, which leads to the following relationship between the two weights:
\begin{equation}
\label{eq:c2c3}
c_3 = c_2(1+\frac{2}{\eta t_d})
\end{equation}
In doing so, the total number of parameters in the model has been reduced. The default parameters of the model are shown in Table \ref{tab:Overview}.

\subsubsection{Verification of the ACC model}
To verify whether the proposed ACC model generates plausible human car-following behaviour, we check the mathematical property of the acceleration function (\ref{eq:optaACC}) and perform a face validation of the ACC model. Several authors have provided basic requirements for plausible car-following models \citep{Treiber2011a,Wilson2011a}. Let $u_{mic}(s,\Delta v, v)$ denote a general class of car-following models where the acceleration is a function of gap $s$, relative speed $\Delta v$ and speed $v$. The basic requirements for car-following models can be summarised with:
\begin{enumerate}
\item The acceleration is an increasing function of the gap to the predecessor $\frac{\partial u_{mic}(s,\Delta v, v)}{\partial s} \geq 0$ and is not influenced by the gap when the predecessor is far in front:  
$ \lim_{s \to \infty}\frac{\partial u_{mic}(s,\Delta v, v)}{\partial s} = 0$.
\item The acceleration is an increasing function of relative speed with respect to the preceding vehicle $\frac{\partial u_{mic}(s,\Delta v, v)}{\partial \Delta v} \geq 0$, and is not influenced by the relative speed at very large gaps  $\lim_{s \to \infty}\frac{\partial u_{mic}(s,\Delta v, v)}{\partial \Delta v} = 0$.
\item The acceleration is a strictly decreasing function of speed $\frac{\partial u_{mic}(s,\Delta v, v)}{\partial v} < 0$, and equals zero when vehicles travel with free speed at very large gaps $ \lim_{s \to \infty} u_{mic}(s,\Delta v, v_0) = 0$.
\end{enumerate}

It can be shown that the proposed optimal ACC control law of Eq. (\ref{eq:optaACC}) satisfies the three basic requirements. 

\begin{figure}
\centering
    \subfigure[]
    {
				\includegraphics[width= 7 cm ]{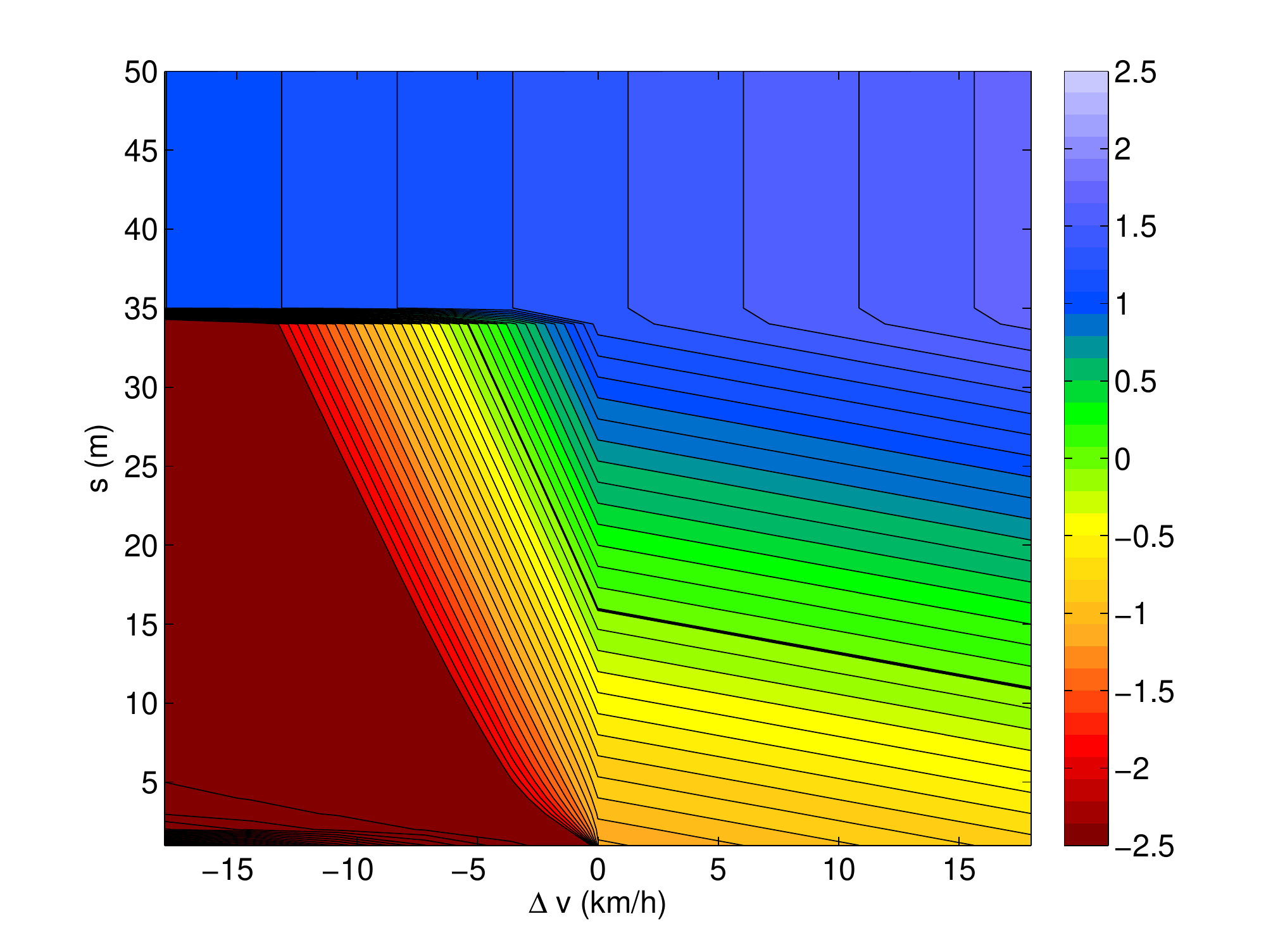}
        \label{fig:ACC_contour_a}
    }
    \subfigure[]
    {
        \includegraphics[width= 7 cm ]{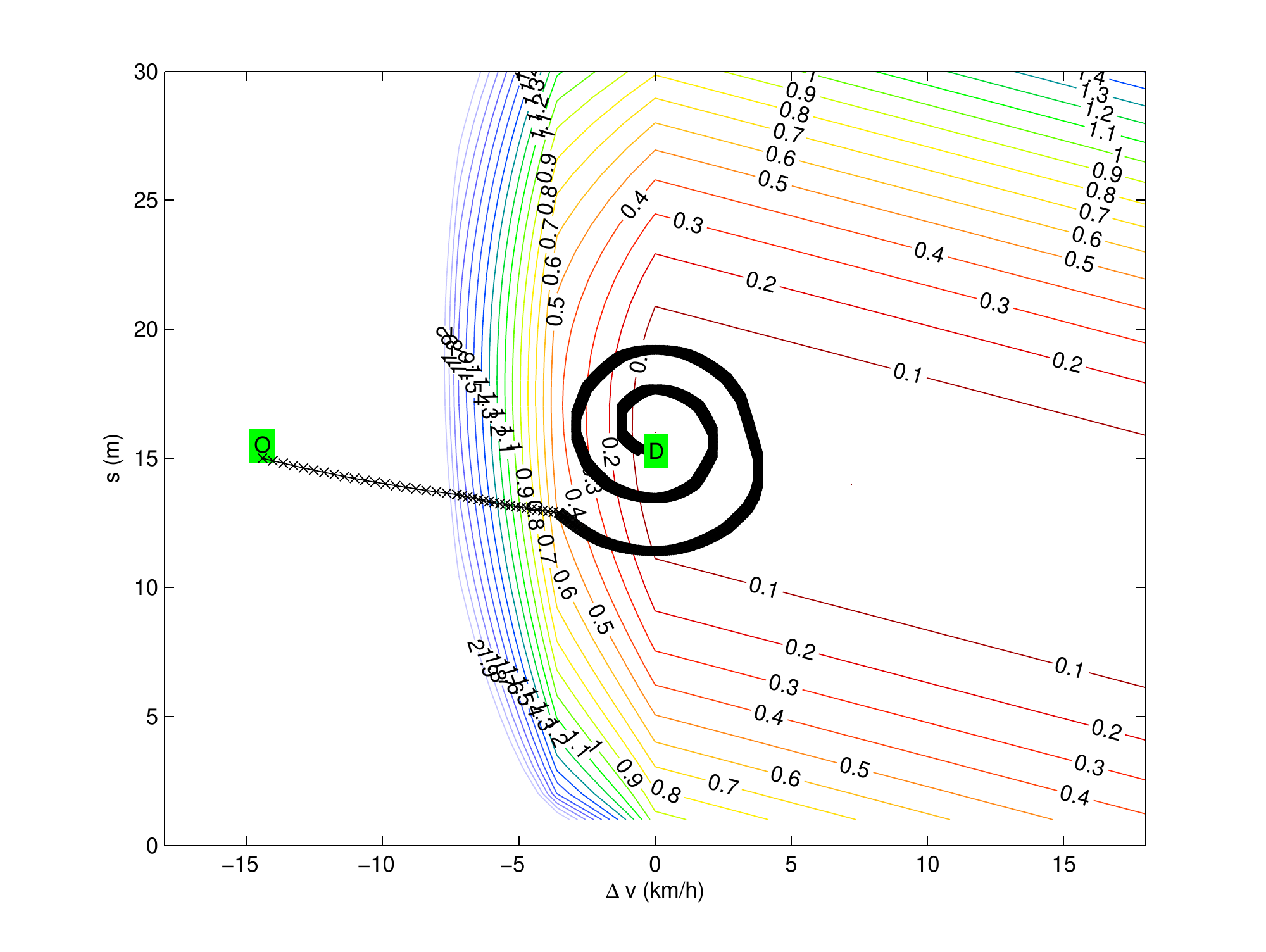}
        \label{fig:ACC_cost_trajectory}
    }
\caption{(a) Contour plot of optimal acceleration when following a  vehicle driving at $54 km/h$; (b) Contour plot of optimal cost with a vehicle trajectory.}.
\label{fig:ISTTT_1}
\end{figure}

Fig. \ref{fig:ACC_contour_a} shows the contour plot of the optimal acceleration for different gaps and relative speeds when following a predecessor driving constantly with a speed of $54 km/h$ using default parameters. Clearly we can see the two regimes of following mode and cruising mode distinguished at the gap of around $35 m$. At cruising mode, the acceleration is above zero, because all the possible speeds (between $36 km/h$ and $72 km/h$) in the contour plot are below the free speed of $120 km/h$. In following mode, the acceleration increases with the increase of headway and relative speed, and consequently decreases with the increase of vehicle speed. The thick line between the green and yellow area shows the neutral line where the accelerations equal zero. Most of the left plane in following mode show a negative acceleration, as a result of the safety cost. This asymmetric property of the optimal acceleration prevents vehicles from driving too close to the leader. 

Fig. \ref{fig:ACC_cost_trajectory} shows how the system evolves from a high cost area to a low cost area of an ACC vehicle following a predecessor driving constantly with a speed of $54 km/h$. The initial state is $s = 15 m$ and $\Delta v = - 14 km/h$ ($v = 68 km/h$), denoted with 'O' in the figure,   using the default parameters. The contour lines show the cost, while the dark star line shows the trajectory of the vehicle, with the optimal acceleration evaluated every $0.25 s$. At the start, the ACC controller incurred safety cost due to approaching the leader and travel efficiency cost due to driving higher than the desired speed of around $47 km/h$. The vehicle starts to decelerate until the relative speed is $0 km/h$. Then it continues to decelerate because driving at $54 km/h$ is still higher than the desired speed, which has changed to around $36 km/h$ (at the gap of $12 m$). As a result, the vehicle will travel with a lower speed and the gap to the predecessor will increase, leading to an increase of the desired speed. The vehicle starts to accelerate when the desired speed is higher than the vehicle speed. The trade-off between the travel efficiency and safety cost will finally lead to the behaviour as shown in the figure, ending with 'D' in the figure after a simulation period of $50 s$.

\begin{table}
\caption{Model parameters}
\label{tab:Overview}
\begin{center}
\begin{tabular}{cl cl cl c}
\hline
Parameter & Physical meaning & Default value & Unit\\
\hline
$v_0$ & free speed &  120 & $km/h$ \\
$c_1$   & weight on safety cost & 0.1 & $s^{-2}$ \\
$c_2$  & weight on efficiency cost & 0.001 & $s^{-2}$\\
$\eta$  & discount factor & 0.25 & $s^{-1}$ \\
$t_d$   & desired time gap & 1.0 & $s$\\
$s_0$ & desired gap at standstill & 1  & $m$\\
$l $  & vehicle length          & 5 & $m$\\
\hline
\end{tabular}
\end{center}
\end{table}

\subsection{Example 2: Cooperative-ACC model}
As a second example, we apply the control framework to design Cooperative-ACC (C-ACC) systems where the controlled vehicle does not only consider its own situation but also the situation of its follower when making control decisions. 
The cooperation mechanism is applied when one C-ACC vehicle is followed by another C-ACC vehicle. In that situation, the two C-ACC vehicles exchange their gaps and relative speeds with each other through V2V communications and they collaborate to minimise a joint cost function, reflecting the situation of both C-ACC vehicles.

\subsubsection{Joint running cost function for C-ACC}
The cooperative behaviour entails minimising a joint cost. Since there is no interaction in cruising mode, we assume that the cooperative behaviour only occurs when both the controlled vehicle and its follower are operating in following mode. Thus we only change the running cost at following mode, which becomes:
\begin{equation}
\label{eq:ACCcost_cooperative}
{\cal L} = c_1 \sum\limits_{j=n}^{n+1}e^{\frac{s_0}{s_j}}\Delta v_j^2 \cdot\Theta(\Delta v_j) + c_2 \sum\limits_{j=n}^{n+1}(v_j - v_d(s_j))^2 + \frac{1}{2}\sum\limits_{j=n}^{n+1}u_j^2
\end{equation}
The running cost function (\ref{eq:ACCcost_cooperative}) shows that in following mode, the cooperative controller aims to minimising the acceleration, safety cost due to approaching the preceding vehicle and efficiency cost due to not driving at desired speed of the C-ACC vehicle and its follower. 

\subsubsection{Optimal control of C-ACC vehicles}
Following solution (\ref{eq:optimalcontrol_cooperative}), we arrive at:
\begin{eqnarray}
\label{eq:opta_cooperative}
u^* &=& \frac{2 c_1e^{\frac{s_0}{s_n}}}{\eta} \left( \Delta v_n  - \frac{s_0\Delta v_n^2}{\eta s_n^2} \right)\cdot\Theta(\Delta v_n) +  \frac{2c_2}{\eta}(1 + \frac{2}{\eta t_d}) \left(v_d(s_n) - v_n \right) \nonumber \\ 
& & {} - \frac{2 c_1e^{\frac{s_0}{s_{n+1}}}}{\eta} \left( \Delta v_{n+1}  - \frac{s_0\Delta v_{n+1}^2}{2\eta s_{n+1}^2} \right)\cdot\Theta(\Delta v_{n+1}) - \frac{2c_2}{ \eta^2 t_d}(v_d(s_{n+1}) - v_{n+1})
\end{eqnarray}
In Eq. (\ref{eq:opta_cooperative}), the optimal acceleration of a C-ACC vehicle $n$ is a function of gap, relative speed and speed of both itself and its follower (vehicle $n+1$). The first two terms in Eq. (\ref{eq:opta_cooperative}) correspond to the non-cooperative ACC model in Eq. (\ref{eq:optaACC}). The third term shows that the C-ACC vehicle will accelerate when its follower is approaching. The fourth term implies that the C-ACC vehicle tends to decelerate when the follower is travelling below the desired speed and tends to accelerate when vice versa. In doing so, the joint cost function (\ref{eq:ACCcost_cooperative}) is optimised. The backward-looking behaviour in the third and fourth term shows how the follower's situation affects the optimal control.

\section{Equilibrium solutions and stability analysis}
In this section, we present the method for analysing ADAS model characteristics, with a focus on equilibrium solution and linear stability analysis. Particularly, we consider a more generalised expression of the optimal controller with cooperative behaviour. The acceleration is expressed as a function of gap, relative speed, and speed of the controlled vehicle $n$ and its follower vehicle $n+1$:
\begin{equation}
\label{eq:generalised_a}
u_n(s_n,\Delta v_n, v_n, s_{n+1},\Delta v_{n+1},v_{n+1}). \nonumber
\end{equation}

\subsection{Equilibrium solutions}
At equilibria in homogeneous traffic, all vehicles travel at the same speed with the same gap and zero acceleration. The equilibrium solutions are derived by the following equation:
\begin{equation}
\label{eq:generalised_FD}
u_n(s_e,0, v_e, s_e,0,v_e) = 0
\end{equation}
which gives a unique equilibrium speed as a function of gap $v_e(s_e)$, or an equilibrium gap as a function of speed $s_e(v_e)$.

\subsection{Linear stability analysis}
The stability analysis framework generalises the classic linear stability analyses approach \citep{Holland1998,Treiber2011a,Wilson2011a} to cooperative systems. Effects on string stability of the cooperative behaviour can be analytically derived. Types of convective instability are classified using signs of signal velocity with a simpler calculation procedure compared to the method of \cite{Ward2011b}.

Let us assume a small deviation $h_n$ and $g_n$ of the $n$th vehicle in the homogeneous platoon from the steady-state gap $s_e$ and speed $v_e$ respectively, then the gap and speed of vehicle $n$ can be written as: 
\begin{equation}
s_n  = s_e + h_n \mbox{ , } v_n  = v_e + g_n
\end{equation}
The first and second order derivatives of $h_n$ give:
\begin{equation}
\label{eq:derivativeh}
\dot{h}_n = \Delta v_n =  g_{n-1}  - g_n \mbox{ , } \ddot{h}_n = u_{n-1} - u_n
\end{equation}
Approximating $u_{n-1}$ and $u_n$ in Eq. (\ref{eq:derivativeh}) around equilibria using Taylor series to the first order arrives at:
\begin{eqnarray}
\label{eq:disturbancetransferfunction}
\ddot{h}_n & = & u_s(h_{n-1} - h_n) + u_{\Delta v}(\dot{h}_{n-1} - \dot{h}_n) + u_v \dot{h}_n \nonumber \\
& & {} + u_{s_b}(h_{n} - h_{n+1}) + u_{\Delta v_b}(\dot{h}_{n} - \dot{h}_{n+1})  + u_{v_b} \dot{h}_{n+1}
\end{eqnarray}
with the coefficients (gradients of acceleration) evaluated at equilibria:
\begin{equation}
u_s = \frac{\partial u_n}{\partial s_n} |_e \mbox{ , } u_{\Delta v} = \frac{\partial u_n}{\partial \Delta v_n} |_e \mbox{ , } u_v = \frac{\partial u_n}{\partial v_n} |_e \mbox{ , } u_{s_b} = \frac{\partial u_n}{\partial s_{n+1}} |_e \mbox{ , } u_{\Delta v _b} = \frac{\partial u_n}{\partial \Delta v_{n+1}} |_e \mbox{ , } u_{v_b} = \frac{\partial u_n}{\partial v_{n + 1} } |_e  \nonumber
\end{equation}

The equilibrium solutions $v_e(s_e)$ restrict the coefficients from being independent from each other.  The acceleration and relative speed along the equilibrium solutions should always be zero. This property leads to the following relationship by approximating acceleration around equilibria with Taylor expansion to the first order:
\begin{equation}
\label{eq:relation_as_av2}
(u_s + u_{s_b})  = -v_e'(s_e) \cdot (u_v + u_{v_b})
\end{equation}

\subsubsection{Local stability criteria}
For local stability, we are primarily interested in a pair of vehicles, where the leader is driving constantly. In this case, Eq. (\ref{eq:disturbancetransferfunction}) will relax to:
\begin{equation}
\label{eq:localstabilitydisturbance}
\ddot{h}_n +(u_{\Delta v} - u_v)\dot{h}_n + u_sh_n = 0
\end{equation}
Equation (\ref{eq:localstabilitydisturbance}) is a harmonic damped oscillator which can be solved using the following \textit{ansatz}:
\begin{equation}
\label{eq:gap_disturbance}
h = h_0 e^{\gamma t}
\end{equation}
where $\gamma = \sigma + i\omega$ ($i = \sqrt{-1}$) is the complex growth rate and $h_0$ reflects the amplitude of the initial disturbance.  We can reformulate the damped oscillator as:
\begin{equation}
\label{eq:oscillator2}
\gamma^2 + \left(u_{\Delta v}-u_v\right) \gamma + u_s= 0
\end{equation}
with solutions
\begin{equation}
\label{eq:oscillatorsolution}
\gamma_{1,2} = \frac{-(u_{\Delta v}-u_v) \pm \sqrt{(u_{\Delta v}-u_v)^2 - 4u_s }}{2}
\end{equation}
Local stability requires both solutions of Eq. \eqref{eq:oscillator2}, $\gamma_1$ and $\gamma_2$, to have negative real parts, which is satisfied by the following condition:
\begin{equation}
\label{eq:localstability}
u_{\Delta v} - u_v > 0
\end{equation}

\subsubsection{String stability criteria}
For string stability, we are interested in how a small disturbance propagates through the increasing index of vehicles. We state the following theorem for string stability of generalised driver assistance system controllers in the form of \eqref{eq:generalised_a}. 

\noindent \textbf{Theorem 1} If $u_v + u_{v_b} < 0$, string stability is guaranteed by the inequality:
\begin{equation}
\label{eq:stringstability_cooperative}
v_e'(s_e)^2 \leq v_e'(s_e) (u_{\Delta v} + u_{\Delta v_b}  -u_{v_b}) + \frac{u_s - u_{s_b}}{2}
\end{equation}

\noindent \textbf{Proof}  The generalised disturbance dynamic equation of (\ref{eq:disturbancetransferfunction}) can be solved using Fourier analysis with the following \textit{ansatz}:
\begin{equation}
\label{eq:Fourier_Ansatz}
h_n = h_0 e^{\gamma t + ink} \mbox{ , } g_n = g_0 e^{\gamma t + ink}
\end{equation}
where $\gamma = \sigma + i\omega$ ( $i = \sqrt{-1}$) is the complex growth rate. The real part $\sigma$ denotes the growth rate of the oscillation amplitude while the imaginary part $\omega$ is the angular frequency from the perspective of the vehicle. The dimensionless wave number $k \in (-\pi, \pi)$ indicates the phase shift of the traffic waves from one vehicle to the next at a given time instant, and the corresponding physical wavelength is $2\pi(s_e + l)/k$  \citep{Treiber2010a}.

To find the limit for string instability, we insert Eq. (\ref{eq:Fourier_Ansatz}) into Eq. (\ref{eq:disturbancetransferfunction}), which yields the following quadratic equation of the eigenvalue $\gamma$:
\begin{equation}
\label{eq:generalised_quadratic_equation}
\gamma^2 + p(k)\gamma + q(k) = 0
\end{equation}
for the complex growth rate $\gamma$ given by 
\begin{equation}
\label{eq:growthrate_gamma}
\gamma_{\pm}(k)  = - \frac{p(k)}{2} \pm \frac{\sqrt{p^2(k)  - 4q(k)}}{2} 
\end{equation}
with coefficients:
\begin{equation}
\label{eq:generalised_p}
p(k) =  u_{\Delta v}(1 - e^{-ik}) - u_{v}  + u_{\Delta v_b} (e^{ik} - 1) - u_{v_b}e^{ik} \mbox{ , }
q(k) = u_{s}(1 - e^{-ik}) + u_{s_b}(e^{ik} - 1)
\end{equation}

For a given \textit{wave number} $k$, only two complex growth rates $\gamma_+$ and $\gamma_-$ are possible and $\mathrm{Re} (\gamma_+) \geq \mathrm{Re} (\gamma_-)$. The model is string stable if $\mathrm{Re}(\gamma) < 0$ for both solutions and for all wave numbers (relative phase shifts) in the range $k \in [-\pi, \pi]$. 

It can be proven that the first instability of time-continuous car-following models without explicit delay always occurs for wave number $k \to 0$ \citep{Wilson2008}. Thus we can expand coefficients of the $p(k)$ and $q(k)$ with Taylor series around $k = 0$:
\begin{equation}
p(k) = p_0 + p_1k + \mathcal{O} (k^2) \mbox{ , } q(k) =  q_1k + q_2 k^2+ \mathcal{O} (k^3)
\end{equation}
with 
\begin{equation}
p_0 = p(0) = - u_v -u_{v_b}  \mbox{ , } p_1 = p'(0) = i(u_{\Delta v} + u_{\Delta v_b} - u_{v_b})  \nonumber
\end{equation}
\begin{equation}
\label{eq:coefficients}
q_1 = q'(0) = i(u_s+u_{s_b}) = iv'_e(s_e) p_0  \mbox{ , } q_2 = \frac{q''(0)}{2} = \frac{u_s - u_{s_b}}{2}
\end{equation}

Expanding root $\gamma_{+}$ around $k=0$ to second order of $k$ and using the Taylor series of square root of $\sqrt{1 - \epsilon} = 1 - \epsilon/2 - \epsilon^2/8 + \mathcal{O} (\epsilon^3)$ gives:
\begin{equation}
\label{eq:gamma2}
\gamma_{+} = -\frac{q_1}{p_0}k + \left( \frac{q_1p_1}{p_0^2} - \frac{q_2}{p_0} - \frac{q_1^2}{p_0^3}   \right) k^2 +  \mathcal{O} (k^3)
\end{equation}
Notice that the first term in Eq. (\ref{eq:gamma2}) is purely imaginary and the second term is a real number. String stability is governed by the sign of the second term. For string stability, it is required that:
\begin{equation}
\label{eq:inequlaity}
\frac{q_1p_1}{p_0^2} - \frac{q_2}{p_0} - \frac{q_1^2}{p_0^3}    \geq 0
\end{equation}

If $u_v + u_{v_b} < 0$, which implies $p_0 >0$, moving the last term in the inequality to the right side and multiply $p_0$ will give:
\begin{equation}
\frac{q_1^2}{p_0^2} \leq \frac{q_1p_1}{p_0} - q_2
\end{equation}

Replacing the coefficients with Eqs. \eqref{eq:coefficients} in the inequality \eqref{eq:inequlaity} and divide by $p_0^2$ will give:
\begin{equation}
v_e'(s_e)^2 \leq v_e'(s_e) (u_{\Delta v} + u_{\Delta v_b}  -u_{v_b}) + \frac{u_s - u_{s_b}}{2}
\end{equation}

\noindent \textbf{Q.E.D.}

For ACC systems that only reacts to the direct predecessor, the string stability criteria relax to: 
\begin{equation}
\label{eq:stringstability_noncooperative}
v_e'(s_e)^2 \leq v_e'(s_e) u_{\Delta v} + \frac{u_s}{2} 
\end{equation}

When comparing Eq. (\ref{eq:stringstability_cooperative}) with Eq. (\ref{eq:stringstability_noncooperative}), we can draw the following analytical criteria for stabilisation effects of cooperative systems. If  a cooperative system keeps the equilibrium speed-gap relationship and the gradients of acceleration $u_s$, $u_{\Delta v}$ and $u_v$ the same as a non-cooperative system, the stabilisation effect of the cooperative behaviour compared to the non-cooperative model, is determined with:
\begin{eqnarray}
\label{eq:stabilisation_cooperative}
v_e'(s_e) (u_{\Delta v_b}  -u_{v_b}) - \frac{ u_{s_b}}{2} > 0 \mbox{, cooperative system is more stable;} \nonumber \\
v_e'(s_e) (u_{\Delta v_b} -u_{v_b}) - \frac{ u_{s_b}}{2} = 0 \mbox{, model stability criteria remains unchanged;}\nonumber \\
v_e'(s_e) (u_{\Delta v_b}  -u_{v_b}) - \frac{u_{s_b}}{2} < 0 \mbox{, cooperative system is more unstable.} 
\end{eqnarray}

\subsubsection{Convective instability}
Several authors discovered that the flow instability in traffic flow are of a convective type \citep{Wilson2011a,Treiber2011a}. Let $Z(x,t)$ denote the spatio-temporal evolution of an initial perturbation $Z(x,0)$. Traffic flow is convectively unstable if it is linearly unstable and if 
\begin{equation}
\label{eq:convectiveinstabilitydefinition}
\lim\limits_{t \to \infty}Z(0,t) = 0
\end{equation}
Intuitively, Eq. (\ref{eq:convectiveinstabilitydefinition}) means that the perturbation will eventually convect out of the system after a sufficient time \citep{Wilson2011a,Treiber2011a}. Otherwise, if traffic flow is linearly unstable but does not satisfy Eq. (\ref{eq:convectiveinstabilitydefinition}), then it is absolutely unstable.


To investigate the limits of convective instability, \cite{Treiber2010a} proposed Fourier transform of a linear response function, which enables one to determine the spatio-temporal evolution of the perturbation $Z(x,t)$. The approach involves finding the wave number corresponding to the maximum growth rate and expanding the complex growth rate around the wave number. After solving a well-defined Gaussian integral, one can obtain the  spatio-temporal evolution of the perturbation as:
\begin{equation}
\label{eq:Fourier_Perturbation5}
Z(x,t) = \mathrm{Re} \frac{Z_0}{\sqrt{-2\pi\gamma''(k^p_0)t}} \exp\left[i(k^p_0x - \omega^p_0 t)\right] \exp\left[ \left( \sigma_0 + \frac{(c_g - \frac{x}{t})^2}{2(i\omega^p_{kk} - \sigma^p_{kk})}\right) t \right] 
\end{equation}
where $k_0^p$ denotes the physical wave number with the maximum growth rate, and is determined by the dimensionless wave number $k_0$: 
\begin{equation}
\label{eq:k0}
k^p_0 = \frac{k_0}{s_e + l} \mbox{ , } k_0 = \arg \max_{k} (\mathrm{Re} \mbox{  } \gamma(k) )
\end{equation}
and
\begin{equation}
\sigma_0 = \mathrm{Re} \gamma(k_0) \mbox{ , } \omega^p_0 = \frac{v_e k_0}{s_e + l} +  \mathrm{Im} \gamma(k^p_0) \mbox{ , } \sigma^p_{kk} = (s_e + l)^2 \mathrm{Re} \gamma''(k_0) \mbox{ , } \omega^p_{kk} = (s_e + l)^2 \mathrm{Im} \gamma''(k_0) \nonumber
\end{equation}
\begin{equation}
\label{eq:50}
c_g = v_e +   (s_e + l) \mathrm{Im} \gamma'(k_0) \mbox{ , } c_p = \frac{\omega_0}{k_0^{p}} = v_e + ( s_e + l) \frac{\mathrm{Im} \gamma(k_0)}{k_0}
\end{equation}
For details, we refer to \cite{Treiber2010a,Treiber2011a}.

In Eq. (\ref{eq:50}), $c_p$ denotes the phase velocity, which is defined by the movement of points of constant phase. It represents the propagation velocity of a single wave.  For human-driven vehicular traffic, the phase velocity $c_p$ is of the order of $- 15 km/h$ in congested traffic  \citep{Treiber2011a}.
$c_g$ is the group velocity,  with which the overall shape of the wave amplitudes propagates through space \citep{Lighthill1965}. More intuitively, the middle of a wave group (or perturbation) propagates with group velocity \citep{Treiber2010a}. The group velocity can be influenced by several waves.

While group velocity represents the propagation of the centre of a wave group, signal velocity $c_s$ is more representative in describing the spatio-temporal dynamics of disturbance in dissipative media like vehicular traffic flow. The signal velocity represents the propagation of waves that neither grow nor decay. It can be calculated using Eq. (\ref{eq:Fourier_Perturbation5}), by considering the growth rate of $Z(x,t)$ along the trajectory of $x = c_st$ and setting it to be zero, which gives:
\begin{equation}
\sigma_0 - \mathrm{Re} \left( \frac{(c_g - c_s)^2}{2\gamma''} \right):= \sigma_0 -   \frac{(c_g - c_s)^2}{2D_2} 
\end{equation}
where $D_2 = -\sigma^p_{kk}\left( 1 + \frac{(\omega_{kk}^p)^2}{(\sigma_{kk}^p)^2} \right)$.
If there is any string instability, we have two signal velocities:
\begin{equation}
\label{eq:signalvelocity}
c_s^{\pm} = v_g \pm \sqrt{2D_2\sigma_0}
\end{equation}
Equation (\ref{eq:signalvelocity}) shows that the perturbed region grows spatially at the constant rate of $2\sqrt{2D_2\sigma_0}$. Convective instability types can be classified as:
\begin{itemize}
\label{eq:convectiveinstabiltyBysignalvelocity}
\item if $c_s^- < 0 < c_s^+$, traffic flow is absolutely string unstable.
\item if $ c_s^+ < 0$, traffic flow is upstream convectively  unstable.
\item if $c_s^- > 0$, traffic flow is downstream convectively unstable.
\end{itemize}

Different from the classification method of using group velocity in \cite{Treiber2010a}, convective instabilities are determined by the signs of signal velocities of disturbance, and the calculation procedure of signal velocity is more approachable to traffic community than that in \cite{Ward2011b}.

\section{ACC and C-ACC model characteristics}
In this section, we use the model analysis framework described in the previous section to examine the characteristics of ACC and C-ACC models. Since there is no interaction with other vehicles in the optimal control input at cruising mode, we emphasize that both local stability and string stability are guaranteed in cruising mode for both the ACC model and the C-ACC model. The stability analyses in the ensuing focus on following mode.

\subsection{Fundamental Diagram}
For the ACC model (\ref{eq:optaACC}), following the equilibrium solutions in the previous section  (when $\Delta v=0$ and $a^*=0$) gives a unique relationship of equilibrium speed and gap:
\begin{equation}
\label{eq:Equilibrium1}
v_e =  \left\{ \begin{array}{cl}
\frac{s_e - s_0}{t_d} &\mbox {if $s_e \leq s_f$} \\
v_0 &\mbox{ if $s_e > s_f$} \\
       \end{array} \right.
\end{equation}

Assuming constant vehicle length $l$ and using the relationship between gap and local density $\rho$:  $\frac{1000}{\rho} = s + l$, we will get the classic triangular fundamental diagram of the steady-state flow-density relationship as:
\begin{equation}
\label{eq:Equilibrium2}
q = \left\{ \begin{array}{cl}
3.6v_0\rho &\mbox { if $\rho \leq \frac{1000}{v_0t_d+s_0+l}$} \\
\frac{1000 - (s_0+l)\rho}{t_d} &\mbox{ if $\rho > \frac{1000}{v_0t_d+s_0+l}$} \\
       \end{array} \right.
\end{equation}
with $q$ denoting traffic flow in the unit of $veh/h$ and $\rho$ in the unit of $veh/km$. 

\begin{figure}
\centering

    \subfigure[]
    {
        \includegraphics[width= 7 cm]{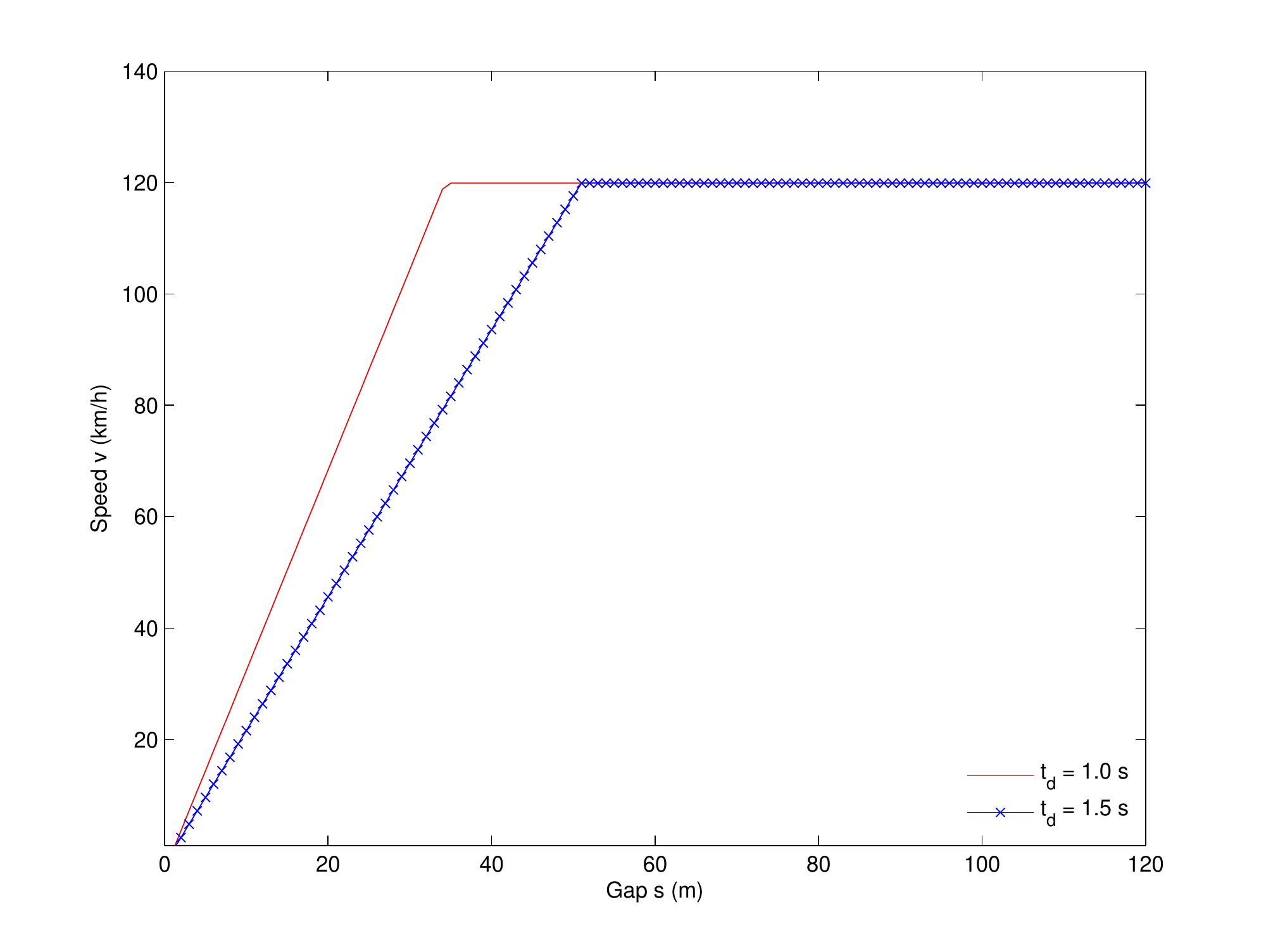}
        \label{fig:ISTTT_FD_speed_gap}
    }
    \subfigure[]
    {
        \includegraphics[width= 7 cm]{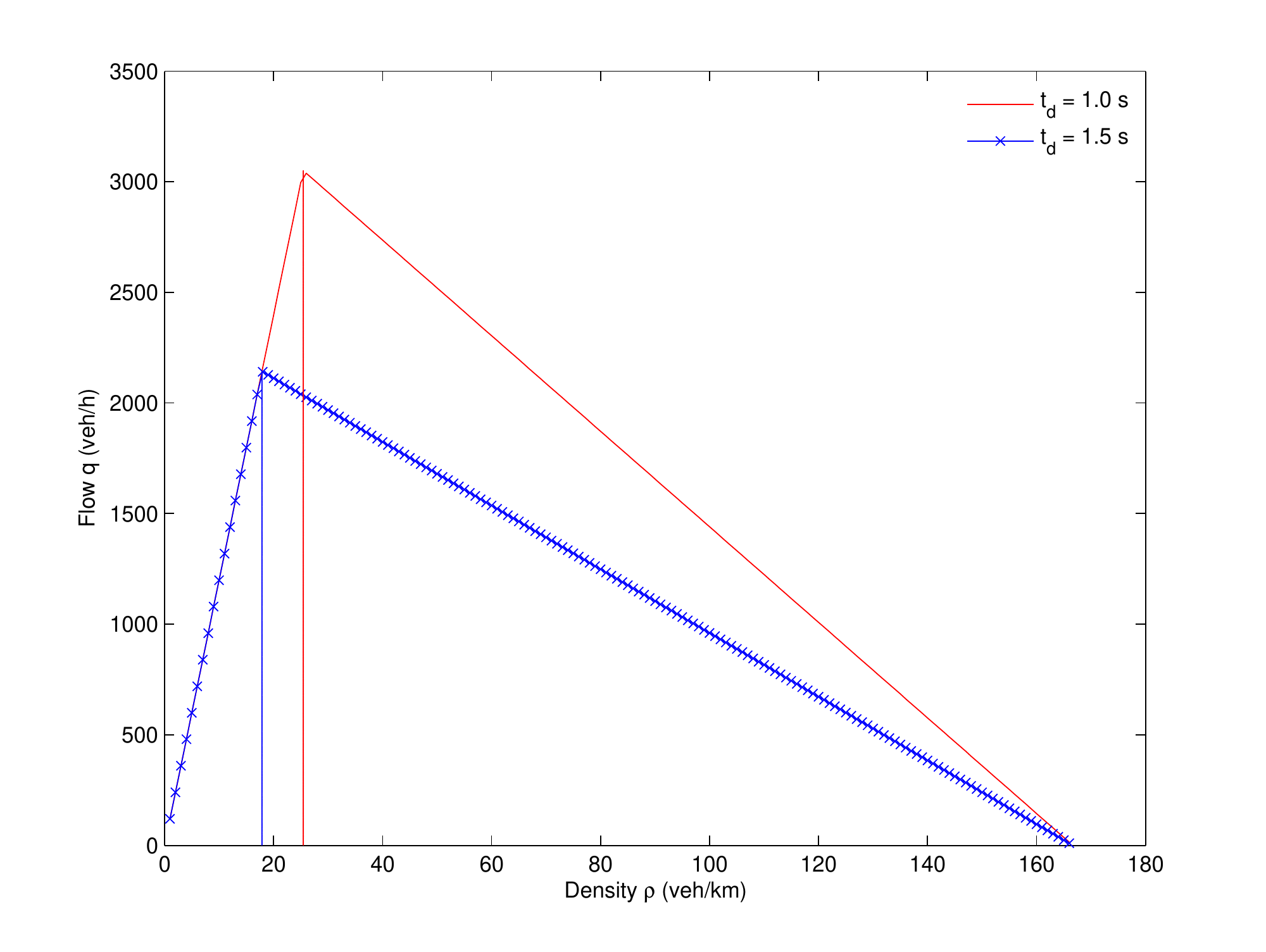}
        \label{fig:ISTTT_FD_flow_density}
    }
\caption{Equilibrium (a) speed-gap relationship and (b) flow-density relationship with $t_d = 1.0 s$ and $t_d = 1.5 s$ and other default parameters in Table \ref{tab:Overview}.}
\label{fig:ISTTT_FD}
\end{figure}

Fig. \ref{fig:ISTTT_FD_speed_gap} shows the steady-state speed-gap relationship and Fig. \ref{fig:ISTTT_FD_flow_density} depicts the equilibrium flow-density relation for two different desired time gaps. The two branches in each of the fundamental diagrams are distinguished by the operating mode of the ACC controller. On the left branch ACC vehicles operate in cruising mode, while at the right branch ACC vehicles operate in following mode. With the default parameter $t_d = 1.0 s$, the resulting flow reaches the capacity of $3050 veh/h$ at a critical density of around $25 veh/km$, while a desired time gap of $1.5 s$ leads to a capacity of $2142 veh/h$ at a critical density of around $ 18 veh/km$. The critical density is determined by the gap threshold $s_f$. The figures shows that the desired time gap has a strong influence on the capacity.

The equilibrium solutions of the C-ACC model are the same as of the new ACC model, and both of them display the fundamental diagram as Eq. (\ref{eq:Equilibrium2}) and Fig. \ref{fig:ISTTT_FD}.

\subsection{Local stability of the ACC model}
Local stability is only interesting for the ACC model. It can be shown with Eq. (\ref{eq:optaACC}) that in following mode $u^*_{\Delta v} > 0$ and $u^*_{v} <0$, thus the local stability condition (\ref{eq:localstability}) is always satisfied. This signifies that the optimal acceleration model of (\ref{eq:optaACC}) is unconditionally local-stable.

\begin{figure}
\centering
\includegraphics[height= 6.5 cm]{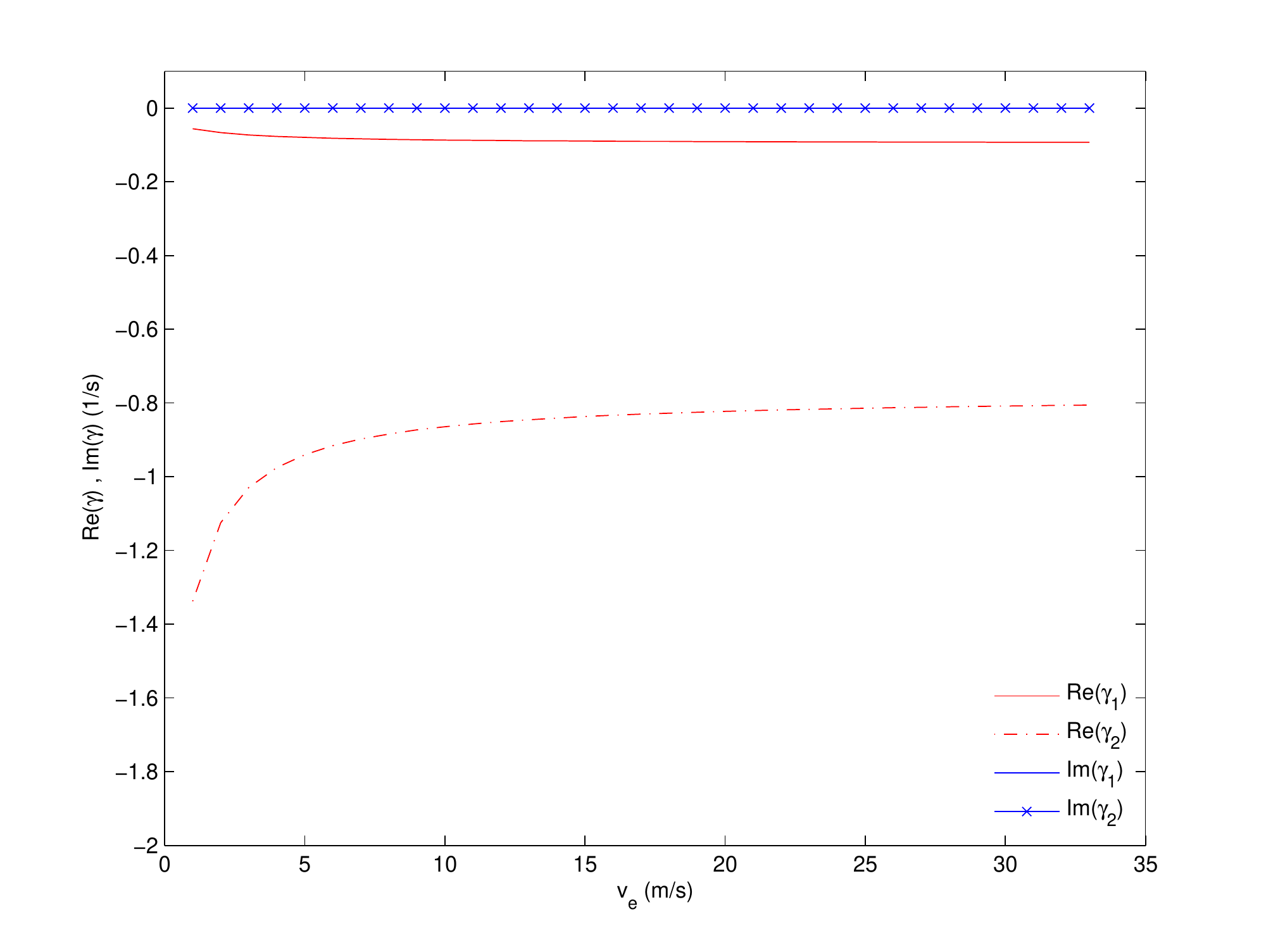}
\caption{Real and imaginary parts of two roots for local stability of the ACC model with default parameters.}
\label{fig:ISTTT_local_stability}
\end{figure}

Fig. \ref{fig:ISTTT_local_stability} shows the two roots of linear growth rate $\gamma_1$ and $\gamma_2$ calculated with solution ({\ref{eq:oscillatorsolution}}). We can clearly see from the figure that the real parts of the two roots are below zero.

\subsection{String stability of the ACC model}
String stability of the proposed ACC model is examined with the linear stability approach. 
\subsubsection{String stability threshold}
To find the string stability threshold, we evaluate the gradients of $u^*$ (\ref{eq:optaACC}) at equilibria and the derivative of equilibrium speed in (\ref{eq:Equilibrium1}) as:
\begin{equation}
\label{eq:gradients_optaACC}
 u^*_s = \frac{2c_2(2+\eta t_d)}{\eta^2 t_d^2} \mbox{ ,  }  u^*_{\Delta v} = \frac{2c_1}{\eta} e^{\frac{s_0}{s_e}} \mbox{ , }  u^*_v = - \frac{2c_2\left(2+ \eta t_d \right)}{\eta^2 t_d} \mbox{ , } v_e'(s_e) = \frac{1}{t_d}
\end{equation}

The stability condition (\ref{eq:stringstability_noncooperative}) gives the following criteria to guarantee string stability:
\begin{equation}
\label{eq:string_stability_criteria_ACC}
\frac{2c_1t_d}{\eta}e^\frac{s_0}{s_e} + c_2(\frac{2}{\eta^2}+\frac{t_d}{\eta}) \geq 1 
\end{equation}
Equation (\ref{eq:string_stability_criteria_ACC}) gives the following properties of model parameters on the string stability:
\begin{itemize}
\item Increasing safety cost weight $c_1$ will stabilise homogeneous flows. Microscopically, a larger $c_1$ leads to a  higher sensitivity to the relative speed and thus a more anticipative driving style, since relative speed reflects future gaps, which is a simple form of anticipation \citep{Treiber2010a}. This explains the stabilisation effects of increasing $c_1$. 
\item Increasing efficiency cost weight $c_2$ will stabilise homogeneous flows. A larger $c_2$ means that the controller has a higher sensitivity to the deviation from the desired speed. Notice that the maximum acceleration is proportional to $c_2$ in Eq. (\ref{eq:c2_max_a}), a larger $c_2$ means a more \textit{responsive} \textit{agile} driving style, which tends to suppress string instabilities \citep{Treiber2010a}. However, physical constraints of vehicles limit the choice of too large $c_2$, i.e. increasing $c_1$ from default value from $0.001 s^{-2}$ to $0.002 s^{-2}$ with other default parameters already changes the maximum acceleration from $2.5 m/s^2$ to $ 5 m/s^2$.
\item Increasing the discount factor $\eta$ will destabilise traffic. Notice that a larger $\eta$ implies a shorter anticipation horizon $\frac{1}{\eta}$, or in other words a more short-sighted driving style. A controller only optimising its immediate situation favours string instability.
\item Increasing the desired time gap $t_d$ will increase the left hand side of the inequality (\ref{eq:string_stability_criteria_ACC}), which implies more stable flow. A larger $t_d$ tends to suppress string instability by following with a larger distance at equilibria.
\end{itemize}

\begin{figure}
    \centering
    \subfigure[]
    {
        \includegraphics[width=7cm]{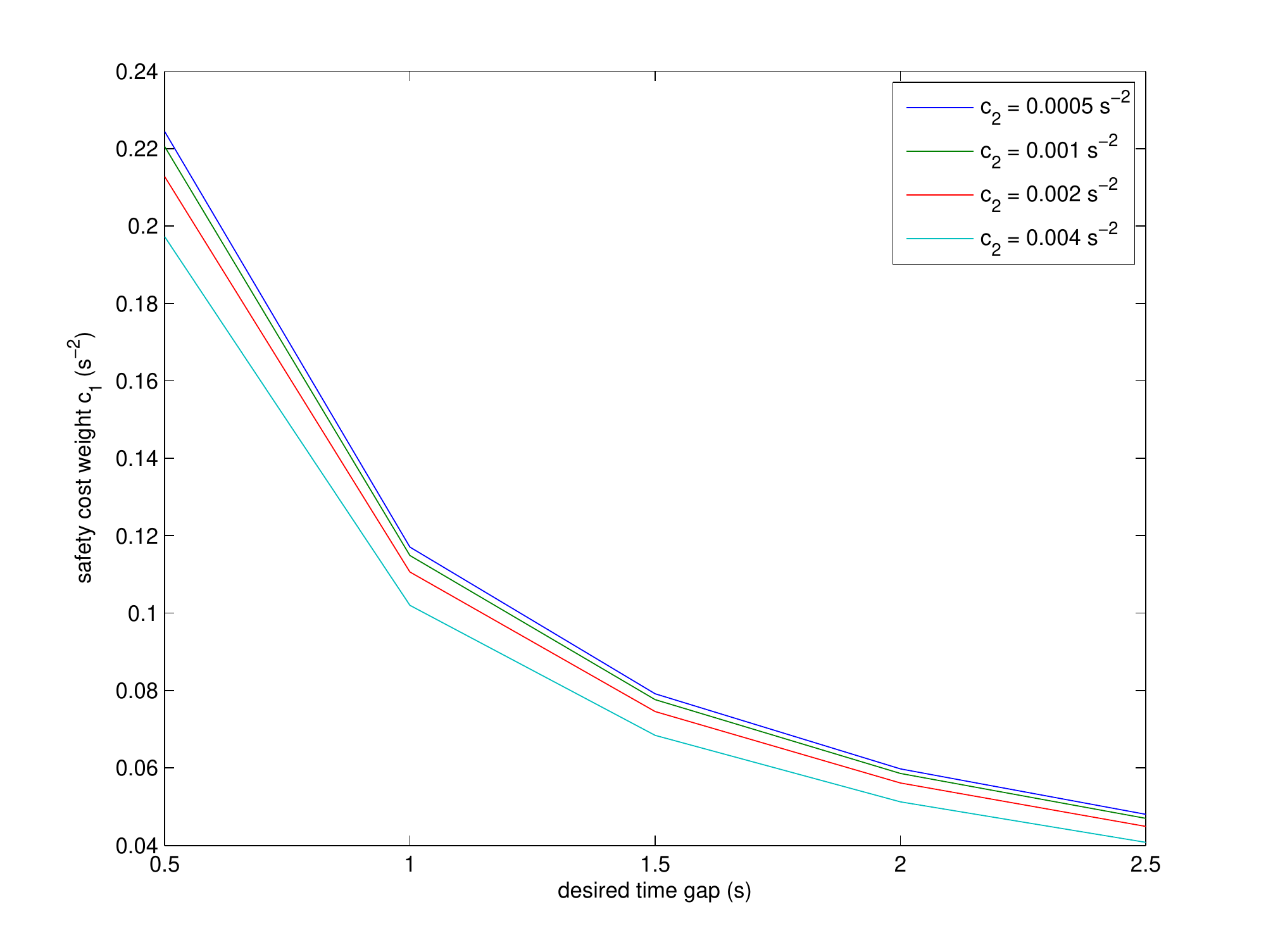}
        \label{fig:first_sub}
    }
    \subfigure[]
    {
        \includegraphics[width=7cm]{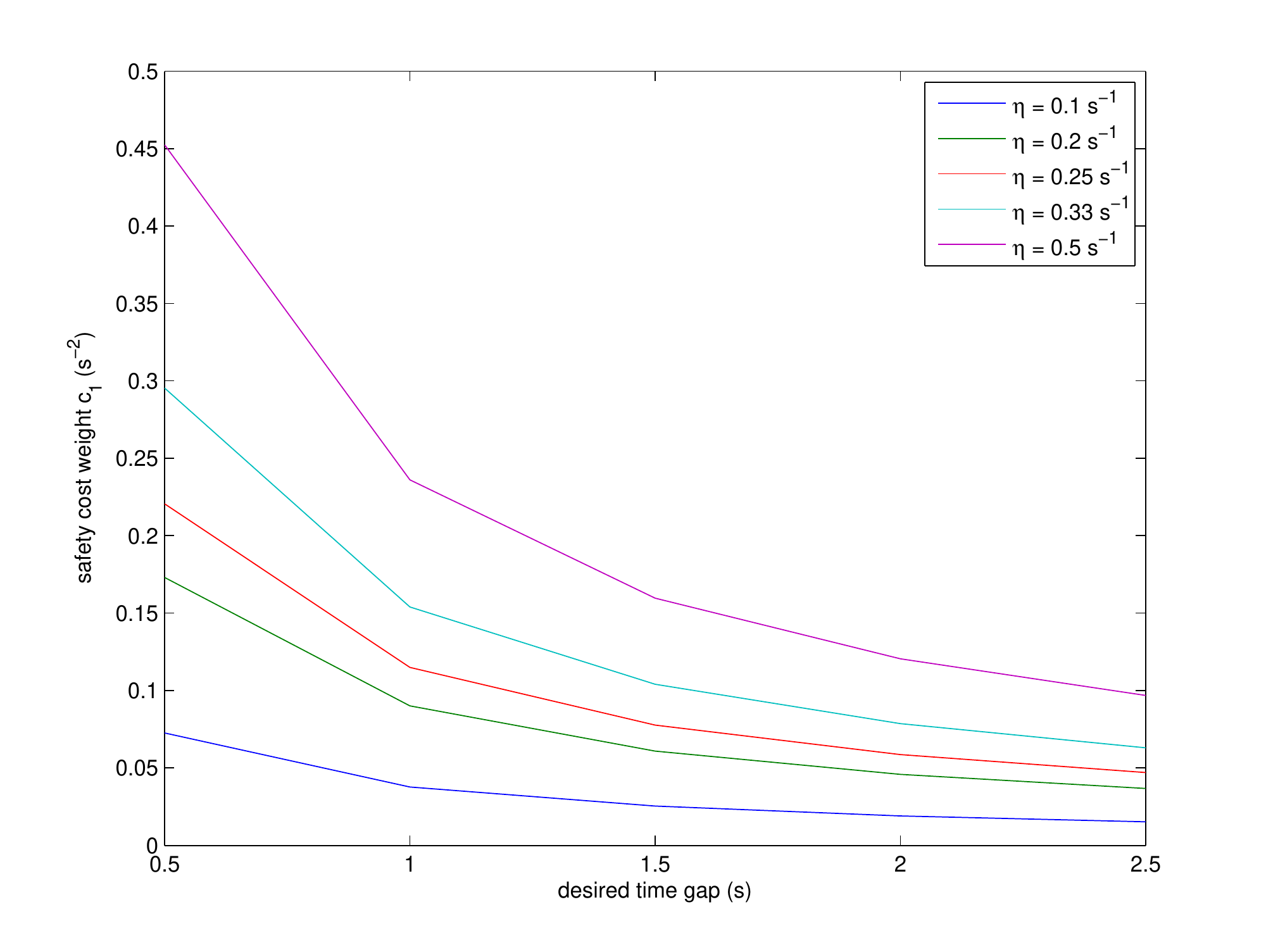}
        \label{fig:second_sub}
    }

    \caption{Stability region in a two-dimensional parameter plane of $c_1$ and $t_d$ with (a) different $c_2$ and (b) different $\eta$,  under equilibrium speed of 72 $km/h$. Other parameters are default values.}
    \label{fig:Stabilityregion_c1_td}
\end{figure}

Fig. \ref{fig:Stabilityregion_c1_td} shows thresholds of stability and instability with different parameters in a two-dimensional parameter plane. The area above the line is string-stable under those parameter settings, while the area below the lines is string-unstable. The stabilisation effects of the parameters are clearly seen.

\subsubsection{Convective instability}
With Eq. (\ref{eq:generalised_p}), the coefficients of the quadratic equation for the complex growth rate $\gamma$ of the ACC model are specified:
\begin{equation}
p(k) = \frac{2c_1}{\eta} e^{\frac{s_0}{s_e}} (1-e^{-ik}) +\frac{2c_2\left(2+ \eta t_d \right)}{\eta^2 t_d} \mbox{ , } q(k)  = \frac{2c_2(2+\eta t_d)}{\eta^2 t_d^2} e^{-ik}
\end{equation}
The first and second order derivatives of $p(k)$ and $q(k)$ can be obtained straightforwardly.

\begin{figure}
\centering
    \subfigure[]
    {
        \includegraphics[width=7cm]{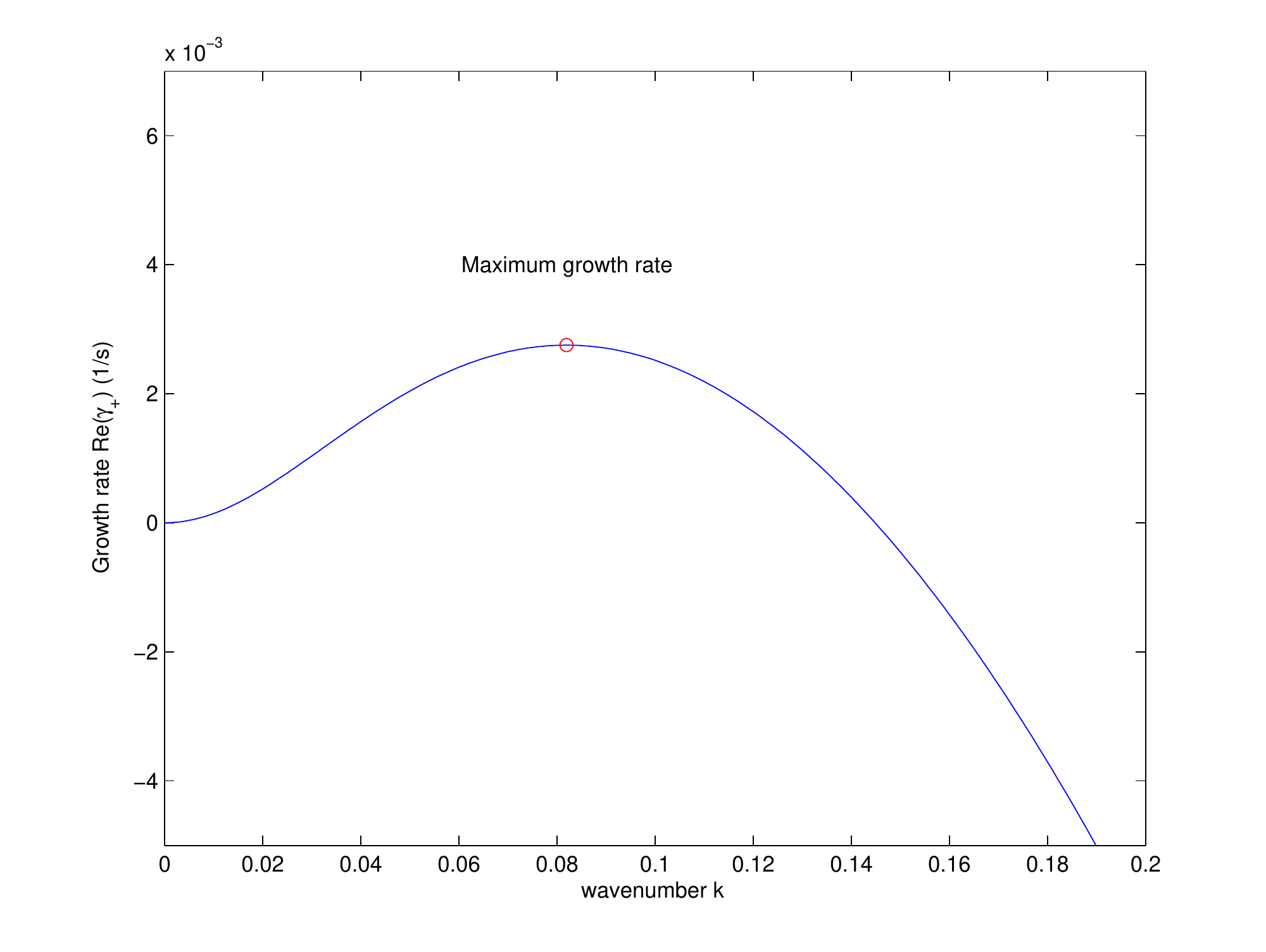}
        \label{fig:ISTTT_ACC_k_sigma}
    }
    \subfigure[]
    {
        \includegraphics[width=7cm]{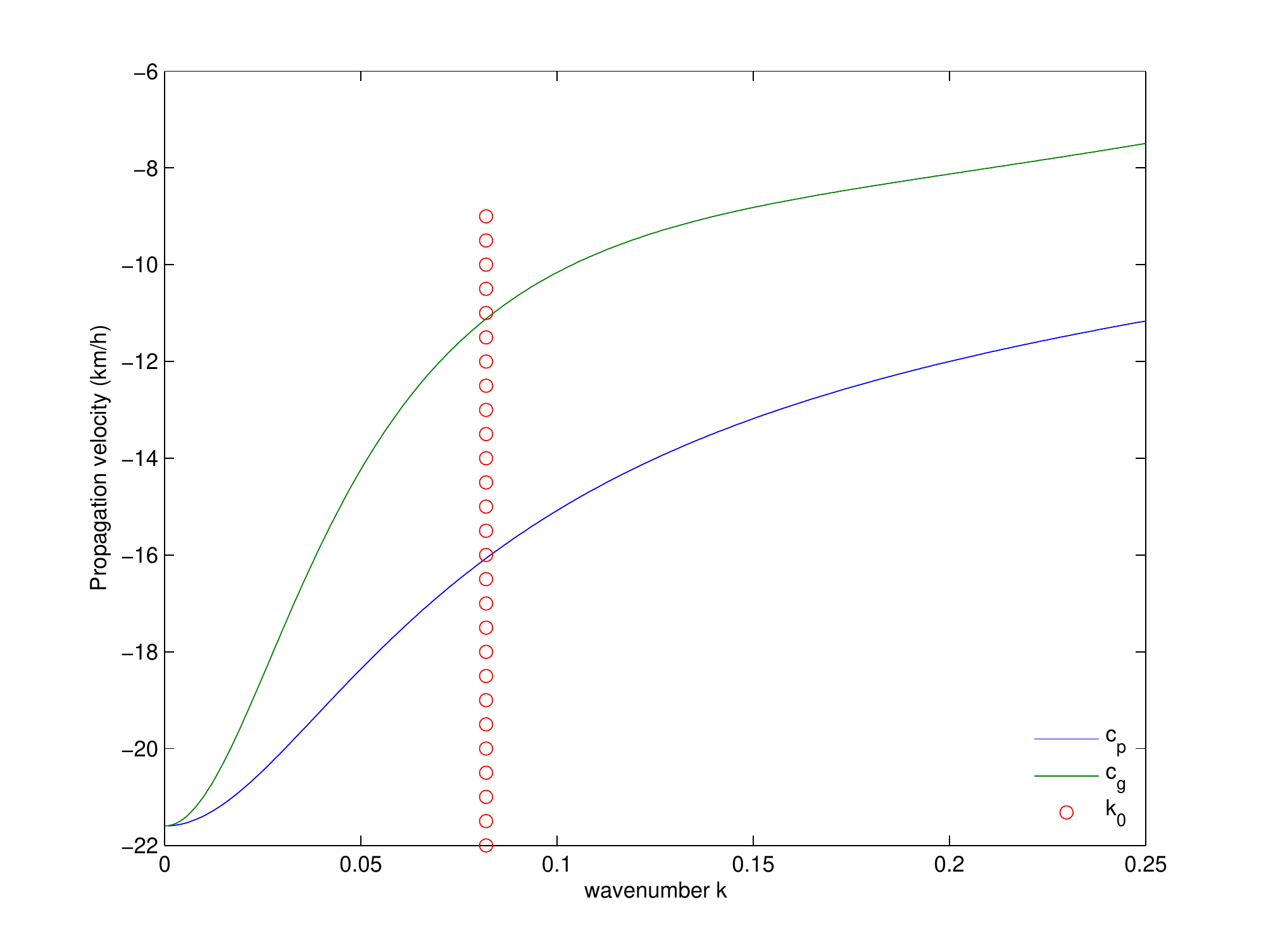}
        \label{fig:ISTTT_ACC_k_vp_vg}
    }
\caption{(a) Growth rate of the more unstable branch $\gamma_+ $ as a function of wave number under $v_e = 54 km/h$ ; (b) phase and group velocity as a function of wave number under $v_e = 54 km/h$ of ACC model with default parameters. }.
\label{fig:ISTTT_ACC_k_sigma_propagationvelocity}
\end{figure}

The linear stability analysis framework enables one to draw the linear growth rate and the propagation velocities of disturbance for the ACC model as a function of wave number under equilibrium speed of 54 $km/h$, as depicted in Fig. \ref{fig:ISTTT_ACC_k_sigma_propagationvelocity}. Numerically, we can find the dimensionless wave number $k_0$ corresponding to the maximum growth rate with the argument (\ref{eq:k0}), which is 0.082 in this case. The physical wavelength is $(s_e+l)2\pi /k_0 \approx 1.5 km$ and the number of vehicles per wave is around $2\pi /k \approx 77$ vehicles. The maximum growth rate is  $0.0028 s^{-1}$ (the red point in the Fig. \ref{fig:ISTTT_ACC_k_sigma}), which is a slow growth implying that it may take some time for an small disturbance grows to traffic breakdown \citep{Treiber2010a}. The phase and group velocity corresponding to this maximum growth rate are  $-16 km/h$ and $-11 km/h$ respectively, with negative sign indicating the propagation direction is against vehicle travelling direction, as depicted in Fig. \ref{fig:ISTTT_ACC_k_vp_vg}.

\begin{figure}
\centering
%
    \subfigure[]
    {
        \includegraphics[width=7cm]{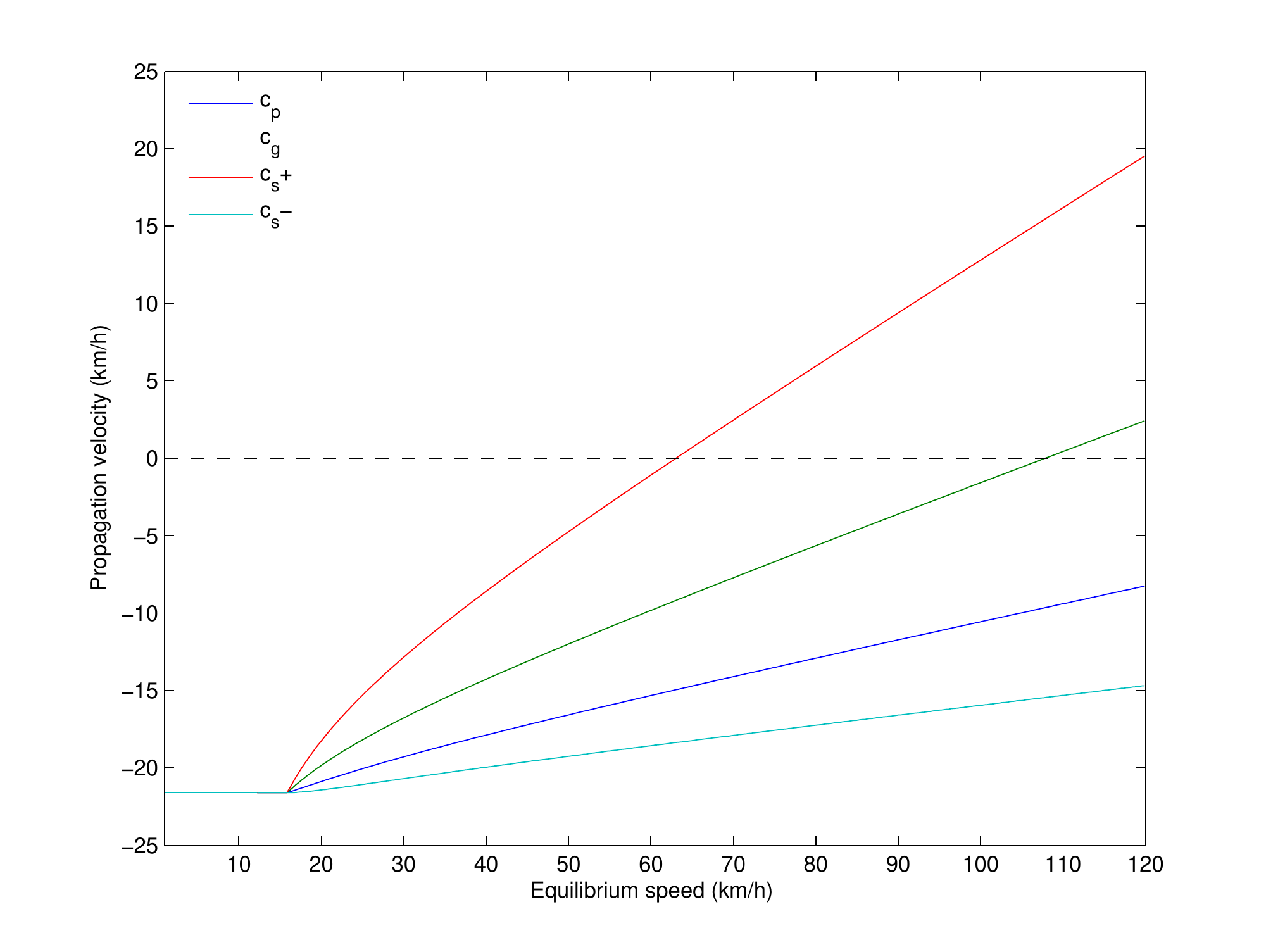}
        \label{fig:vg_ve}
    }
    \subfigure[]
    {
        \includegraphics[width=7cm]{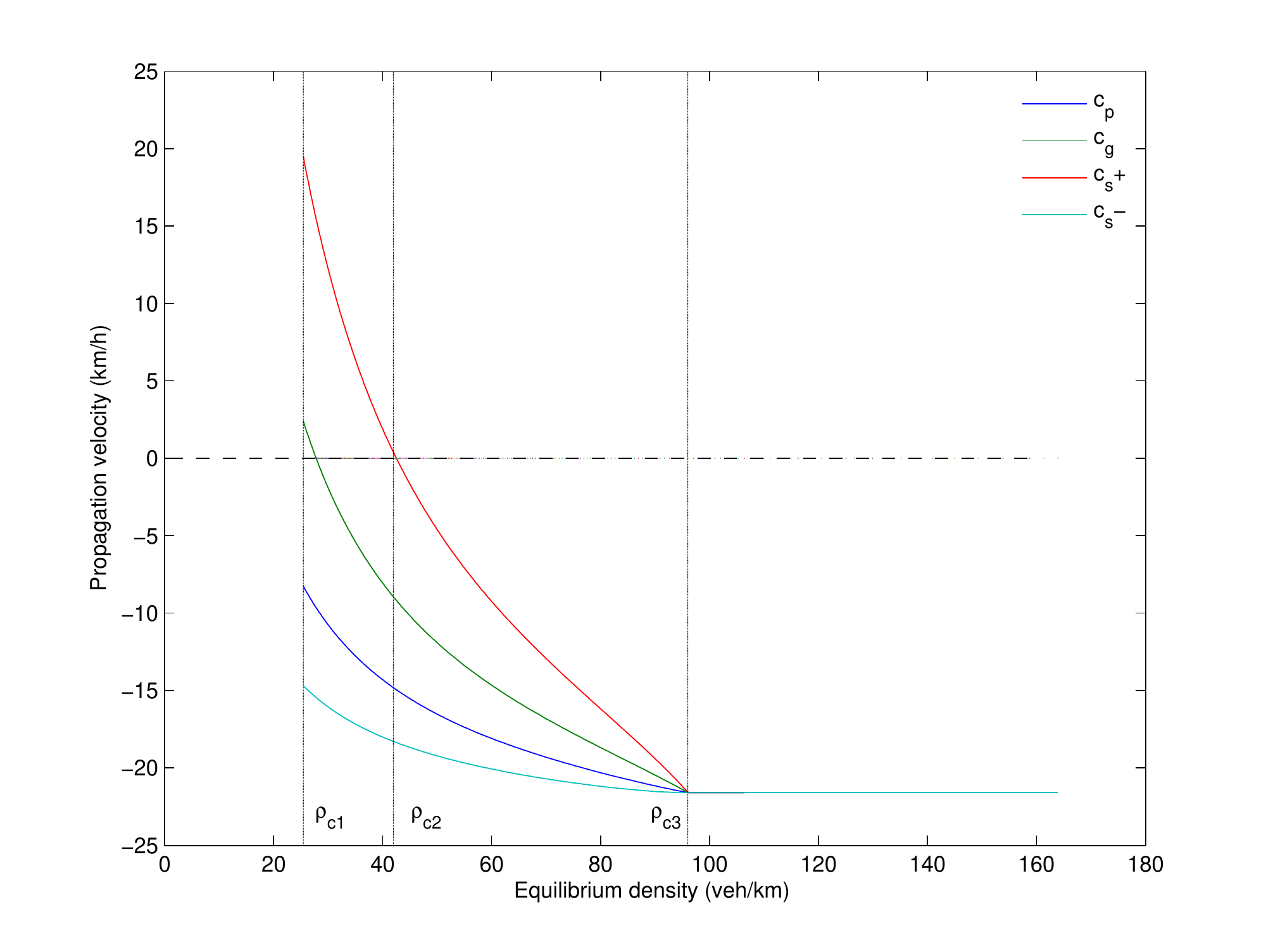}
        \label{fig:vg_rho_e}
    }
    \\
    \subfigure[]
        {
            \includegraphics[width=7cm]{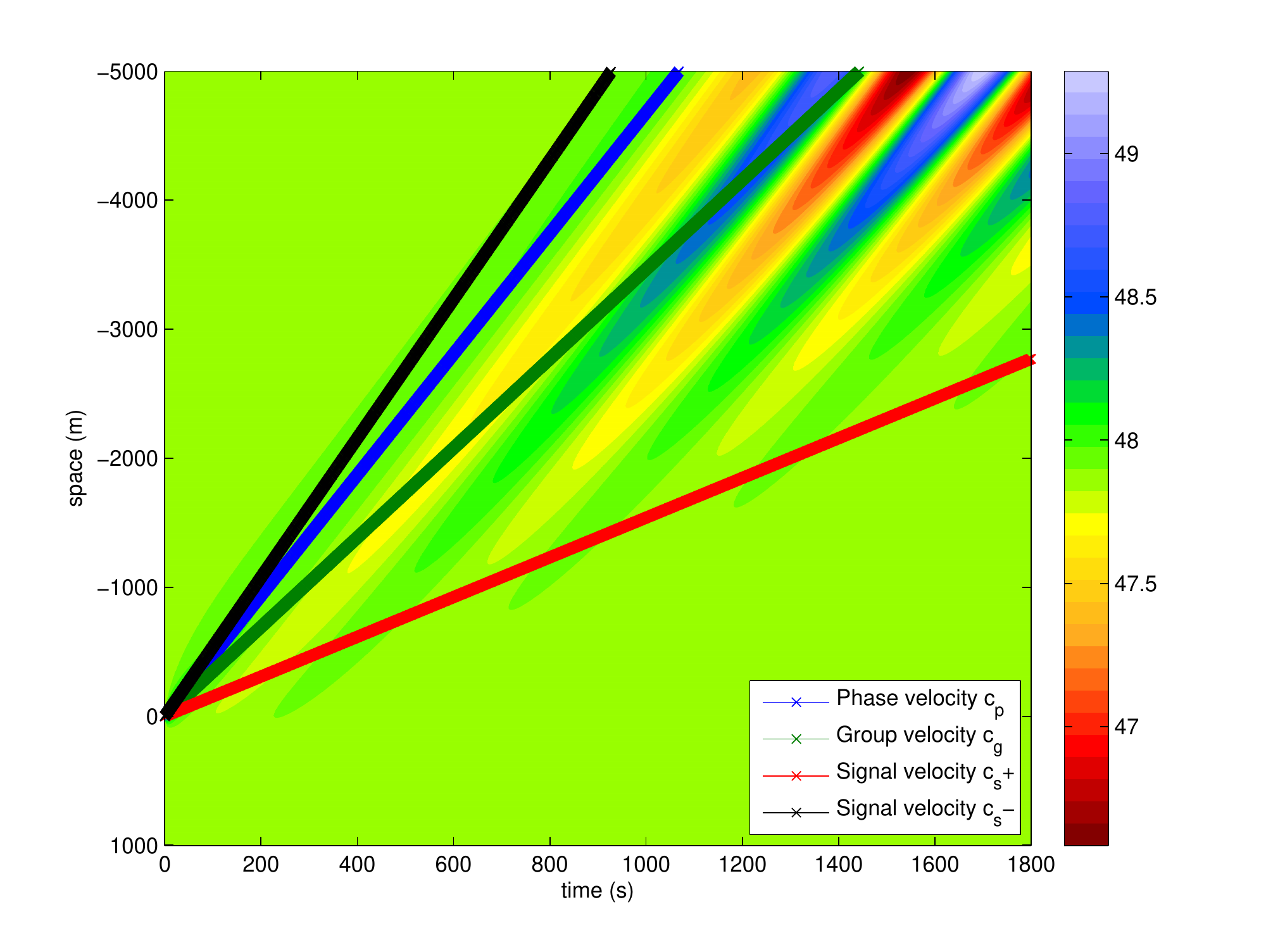}
            \label{fig:evolutionat48}
        }
    \subfigure[]
    {
        \includegraphics[width=7cm]{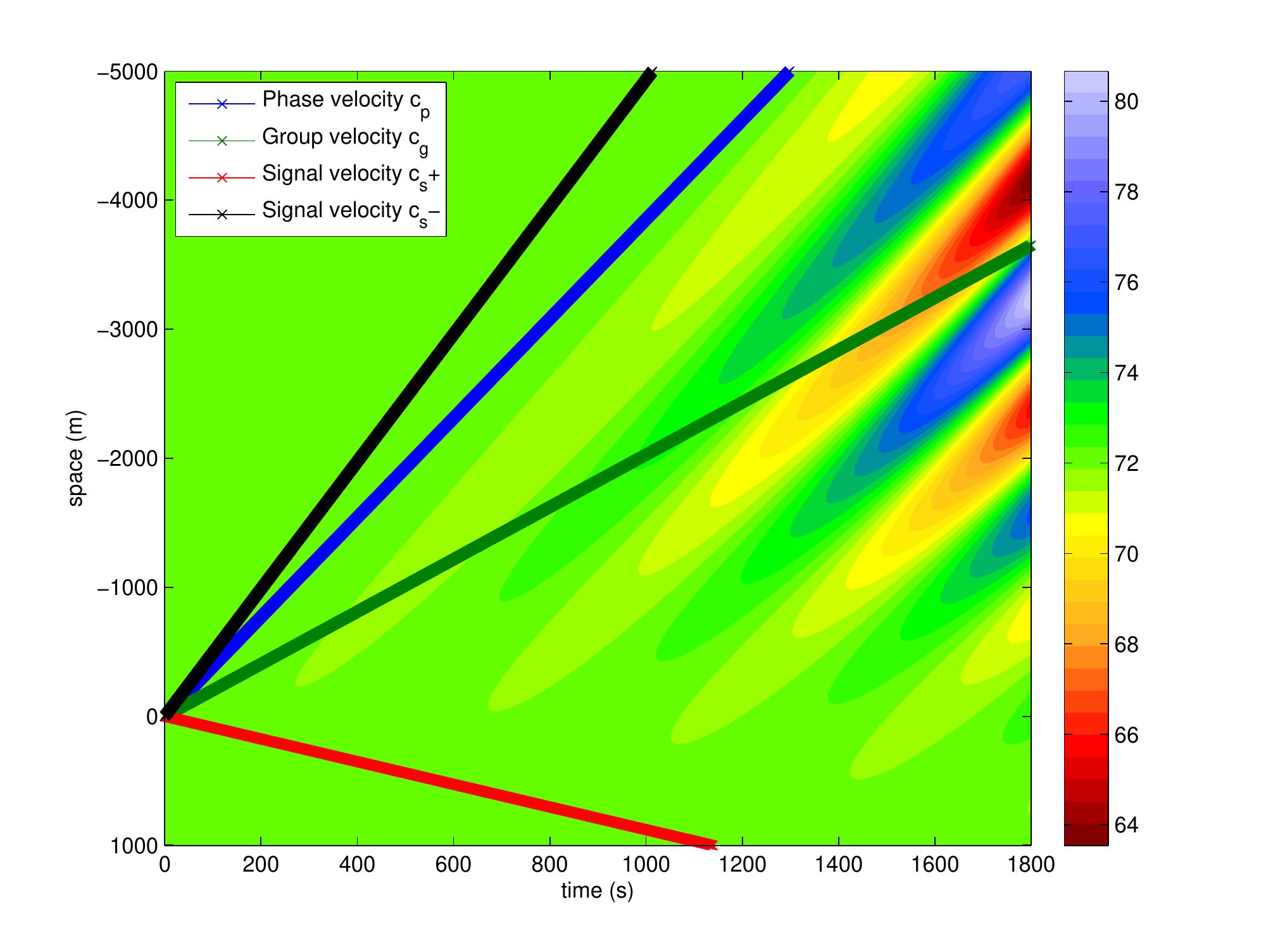}
        \label{fig:evolutionat72}
    }
\caption{(a) Phase, group, signal velocities as a function of equilibrium speed and (b) phase, group, signal velocities as a function of equilibrium density and spatio-temporal evolution of initial disturbance at equilibrium speed of (c) 48 $km/h$ and (d) 72 $km/h$ of ACC model with default parameters. Driving direction in (c) and (d) is from top to down.}.
\label{fig:ISTTT_XTevolution_ve48_72}
\end{figure}

Fig. \ref{fig:vg_ve} and \ref{fig:vg_rho_e} show the phase, group and signal velocities as a function of equilibrium speed and density respectively. Since traffic is always string stable in cruising mode, traffic flow is always stable below the critical density of $\rho_{c1} = 1000/(s_f + l) \approx 25 veh/km$. As long as the density is higher than the critical density $\rho_{c1}$, traffic becomes absolutely unstable $c_{s+} > 0$ and $c_{s-}<0$, with disturbances travelling both upstream and downstream. When the density increases to another critical density $\rho_{c2} \approx 42 veh/km $, the traffic  becomes convectively upstream unstable, with disturbances travelling upstream only. When the density increases further to above another critical density $\rho_{c3} \approx 96 veh/km$, the traffic becomes stable again, which is the so-called \textit{restabilisation} effect \citep{Treiber2010a}. With the default parameters, the ACC model displays absolute and convective upstream instability, which is different from human drivers \citep{Treiber2010a,Wilson2011a}. 

Fig. \ref{fig:evolutionat48} and \ref{fig:evolutionat72} show the spatio-temporal evolution of the system using the analytical disturbance function of \ref{eq:Fourier_Perturbation5} with different equilibrium speeds of $48 km/h$ (density of $52 veh/km$) and $72 km/h$ (density of $38 veh/km$). We can clearly see from the figure that: 
\begin{itemize}
\item at equilibrium speed of $48 km/h$, the initial disturbance travels upstream, while at equilibrium speed of $72 km/h$, disturbance travels both upstream and downstream.
\item  absolute instability grows faster in amplitude, which can be see from the ranges of the speeds contour plots.
\item the centre of the disturbance travels with group velocity and each signal wave travels with phase velocity.
\item two signal velocities limit the region of disturbance in the spatio-temporal plane.
\end{itemize}

When choosing different parameters, one can get different stability characteristics of the model. Fig. \ref{fig:ACC_Stability_Diagram} shows the one dimensional parameter safety cost weight $c_1$ and the resulting stability at different equilibrium speeds at following mode with other default parameters. If we increase $c_1$ to a slightly higher value than the default one, traffic will become convectively upstream stable and stable in following mode, which is similar to human-driven vehicular traffic. When choosing $c_1$ higher than $0.12 s^{-2}$, the traffic is always stable, while $c_1$ lower than $0.06 s^{-2}$ leads to co-existence of convective downstream, absolute and convective upstream instability in the congested branch of the fundamental diagram.


\subsection{Destabilisation effect of the C-ACC model}
The local stability is no longer of interest for the C-ACC controller, since we will consider at least three vehicles in the analysis. For the optimal control of C-ACC controller (\ref{eq:opta_cooperative}), the gradients are given:
\begin{eqnarray}
\label{eq:gradients_optaCACC}
u^*_{s_b} = -\frac{2c_2(1+ \eta t_d)}{\eta^2 t_d^2} 
\mbox{ , }  u^*_{\Delta v_b} =  -\frac{2c_1e^{\frac{s_0}{s_{e}}}}{\eta}
\mbox{ , }  u^*_{v_b}= \frac{2c_2}{\eta^2 t_d}
\end{eqnarray}
while $u^*_s$, $u^*_{\Delta v}$, $u^*_v$ and $v_e'(s_e)$ remain the same as in Eq. (\ref{eq:gradients_optaACC}).

Since $u^*_v + u^*_{v_b} =  < 0 $, condition \eqref{eq:stringstability_cooperative} gives the following criteria for string stability of C-ACC controller:
\begin{equation}
\label{eq:stringstability_criteria_CACC}
\frac{c_2}{\eta^2} (1+\eta t_d) \geq 1
\end{equation}

The stabilisation effect of the C-ACC controller with reference to the ACC controller is governed by (\ref{eq:stabilisation_cooperative}). With the virtue of the gradients in Eq. (\ref{eq:gradients_optaCACC}) and the analytical criteria for the stabilisation effect of cooperative systems (\ref{eq:stabilisation_cooperative}), we found that:
\begin{itemize}
\item $u^*_{s_b} < 0 $, which stabilises traffic. 
\item $u^*_{\Delta v_b} < 0 $, which destabilises traffic.
\item $u^*_{v_b} > 0 $, which destabilises traffic.
\end{itemize}
The total stabilisation effect $v_e'(s_e)\left( u^*_{\Delta v_b} -  u^*_{v_b} \right) - \frac{u^*_s}{2} = -\frac{2c_1}{\eta t_d}e^{\frac{s_0}{s_e}} - \frac{c_2}{\eta^2 t_d^2} <0$, which implies that the C-ACC controller destabilises homogeneous traffic flow compared to the ACC controller. With default parameters, $|u^*_{\Delta v_b}|$ is much larger $|u^*_{s_b}|$ and $|u^*_{v_b}|$, thus this term deteriorates string stability most. 

To classify the convective instability, we need to specify the coefficients of the quadratic equation \eqref{eq:generalised_quadratic_equation} as:
\begin{equation}
\label{eq:p_cooperative}
p(k) = u^*_{\Delta v}(1- e^{-ik} ) + u^*_{\Delta v_b} ( e^{ik} - 1) - u^*_v - u^*_{v_b} e^{ik} \mbox{ , } q(k) = u^*_s(1 - e^{-ik}) + u^*_{s_b} (e ^{ik} - 1)
\end{equation}
The first and second order derivatives of $p(k)$ and $q(k)$ can be obtained straightforwardly.

\begin{figure}
\centering
    \subfigure[]
    {
        \includegraphics[width= 7cm]{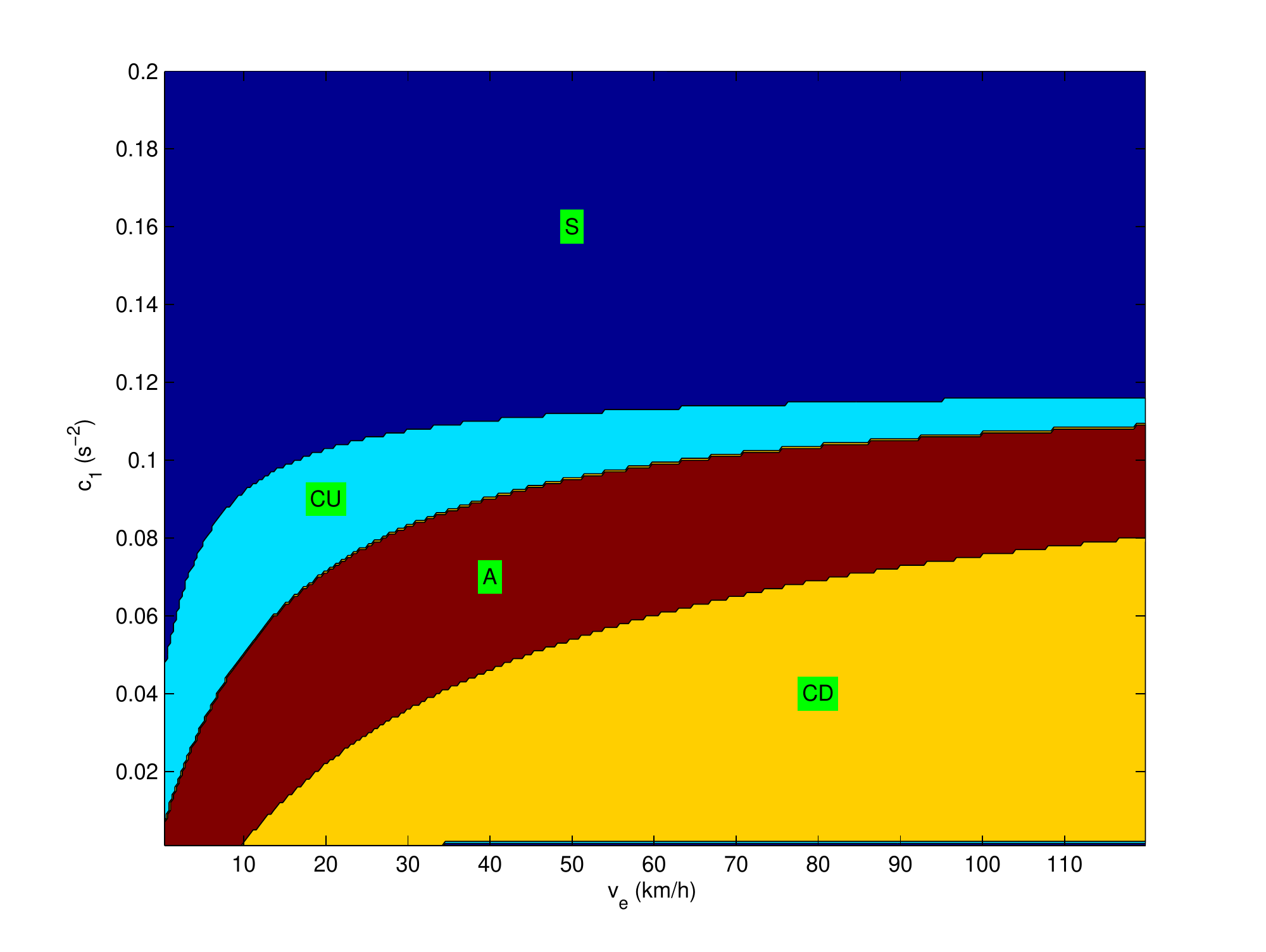}
        \label{fig:ACC_Stability_Diagram}
    }
    \subfigure[]
    {
				\includegraphics[width= 7cm]{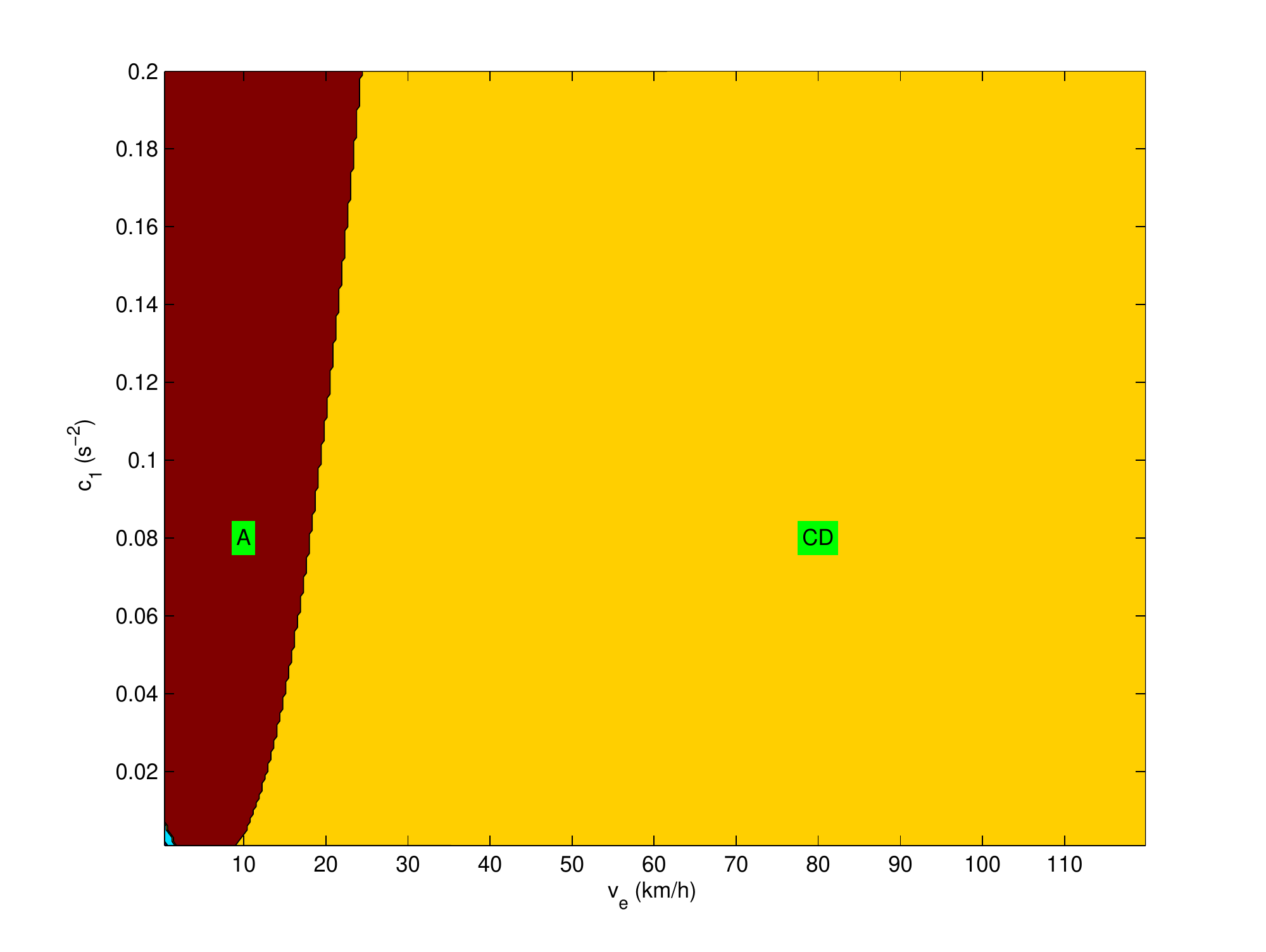}
        \label{fig:C-ACC_Stability_Diagram}
    }
\caption{Stability plot with safety cost weight $c_1$ and equilibrium speed of (a) ACC model; (b) C-ACC model. S: Stable region; U: region with convective Upstream instability; A: region with Absolute instability; D: region with convective Downstream instability.}.
\label{fig:ISTTT_CACCStabDiag_c1_ve}
\end{figure}

The linear stability analysis framework enables us to calculate signal velocity at different equilibrium speeds and different parameter settings. Fig. \ref{fig:C-ACC_Stability_Diagram} shows the resulting stability/instability types of one dimensional parameters. It is quite clear that the C-ACC controller (\ref{eq:opta_cooperative}) is much more unstable compared to the ACC controller in Fig. \ref{fig:ACC_Stability_Diagram}. Homogeneous traffic flow is always unstable in following mode, and the instability is of absolute and convective downstream type.

As a last remark, the analytical stabilisation effects of (\ref{eq:stabilisation_cooperative}) give guidance on how to improve the stability of C-ACC systems. If one can decrease $u^*_{s_b}$ and $u^*_{v_b}$ while  increasing $u^*_{\Delta v_b}$, the string stability of the C-ACC controller will be enhanced. This can be achieved by choosing a different joint cost function.

\section{Conclusion}
We have proposed a control framework to model driver support and cooperative systems, under which the supported driving process is recast into a receding horizon optimisation problem. The control framework is generic such that different objective functions can be minimised with flexible system state specifications.

To show the applicability of the model, we proposed an optimal ACC and an optimal C-ACC controller. The ACC controller has an explicit safety mechanism to prevent collisions and generates plausible car following behaviour.

To gain insights into the macroscopic behaviour of the driver assistance and cooperative systems, we extended the linear stability analysis approach to a cooperative driving environment and derived the string stability criteria for cooperative systems. We analytically quantified the stabilisation effect of cooperative systems with reference to non-cooperative systems. 

We found that the proposed ACC model is unconditionally local-stable, and with careful choice of parameters, the ACC model only displays convective upstream instability at following mode, which is similar to human car-following models. Increasing safety cost weight, efficiency cost weight and desired time gap will stabilise traffic, while increasing the cost discount factor (decreasing the anticipation horizon)  will destabilise traffic. The C-ACC model which optimises the situation of both the controlled vehicle and its follower results in convective downstream and absolute instability type, as opposed to the convective upstream instability type observed in human-driven traffic and the ACC model.

The control framework and analytical results provide guidance in developing controllers for driver assistance systems and give insights into the influence of ACC and C-ACC systems on traffic flow operations.

Future research is directed to investigation of the flow characteristics with different penetration rate of driver assistance systems and the collective behaviour of platoon controller where multi-vehicle are controlled simultaneously.

\section*{Acknowledgement}
The research presented in this paper is part of the research project {\textquotedblleft Sustainability Perspectives of Cooperative Systems \textquotedblright} sponsored by SHELL company.









\bibliographystyle{elsarticle-harv}


\end{document}